\documentclass[aps,prb,twocolumn,superscriptaddress]{revtex4-2}
\usepackage{graphicx}
\usepackage{latexsym}
\usepackage{amssymb}
\usepackage{amsmath}
\usepackage{amsfonts}
\usepackage{upgreek}
\usepackage{bm,dsfont}
\usepackage{multirow}
\usepackage{tikz-cd}
\usepackage{color}
\usepackage{hyperref}
\hypersetup{
colorlinks = true,
linkcolor = [rgb]{0.70,0.13,0.13},
citecolor = [rgb]{0.13,0.55,0.13},
urlcolor = [rgb]{0.25, 0.41, 0.88}}

\newcommand{\dd}{\mathrm{d}}
\newcommand{\ii}{\mathrm{i}}

\newcommand{\U}{\mathrm{U}}
\newcommand{\SU}{\mathrm{SU}}

\newcommand{\SO}{\mathrm{SO}}
\newcommand{\Sp}{\mathrm{Sp}}
\newcommand{\Spin}{\mathrm{Spin}}
\newcommand{\g}{\mathfrak{g}}

\renewcommand{\u}{\mathfrak{u}}
\newcommand{\su}{\mathfrak{su}}

\newcommand{\so}{\mathfrak{so}}

\newcommand{\dsZ}{\mathbb{Z}}

\newcommand{\scM}{\mathcal{M}}

\newcommand{\Tr}{\operatorname{Tr}}

\newcommand{\mat}[1]{\left[\begin{matrix}#1\end{matrix}\right]}
\newcommand{\smat}[1]{\left[\begin{smallmatrix}#1\end{smallmatrix}\right]}
\newcommand{\eq}[1]{\begin{equation}#1\end{equation}}
\newcommand{\eqs}[1]{\begin{equation}\begin{split}#1\end{split}\end{equation}}
\newcommand{\eqnref}[1]{Eq.\,\eqref{#1}}
\newcommand{\figref}[1]{Fig.\,\ref{#1}}

\newcommand{\hookuparrow}{\mathrel{\rotatebox[origin=c]{90}{$\hookrightarrow$}}}

\begin{document}

\title{Deconfined Quantum Criticality among Grand Unified Theories}

\author{Yi-Zhuang You}
\affiliation{Department of Physics, University of California, San Diego, CA 92093, USA}
\author{Juven Wang}
\affiliation{Center of Mathematical Sciences and Applications, Harvard University,  Cambridge, MA 02138, USA}

\begin{abstract}
This article overviews the recent developments in applying the idea of deconfined quantum criticality in condensed matter physics to understand quantum phase transitions among grand unified theories in high energy physics in the 4-dimensional spacetime. 
In particular, 
dictated by a mod 2 class 
nonperturbative global mixed gauge-gravitational anomaly,
there can be 
a gapless deconfined quantum critical region between Georgi-Glashow and Pati-Salam models
---
not only the Standard Model occurs as a neighbor phase,
but also beyond the Standard Model phenomena (such 
as deconfined dark gauge force and 
fractionalized/fragmentary fermionic partons) emerge near the critical region.
An invited contribution to Professor Chen-Ning Yang Centenary Festschrift.
\end{abstract}
\maketitle

\section{Introduction}

\begin{figure}[!htbp]
\begin{center}
\includegraphics[width=0.95\columnwidth]{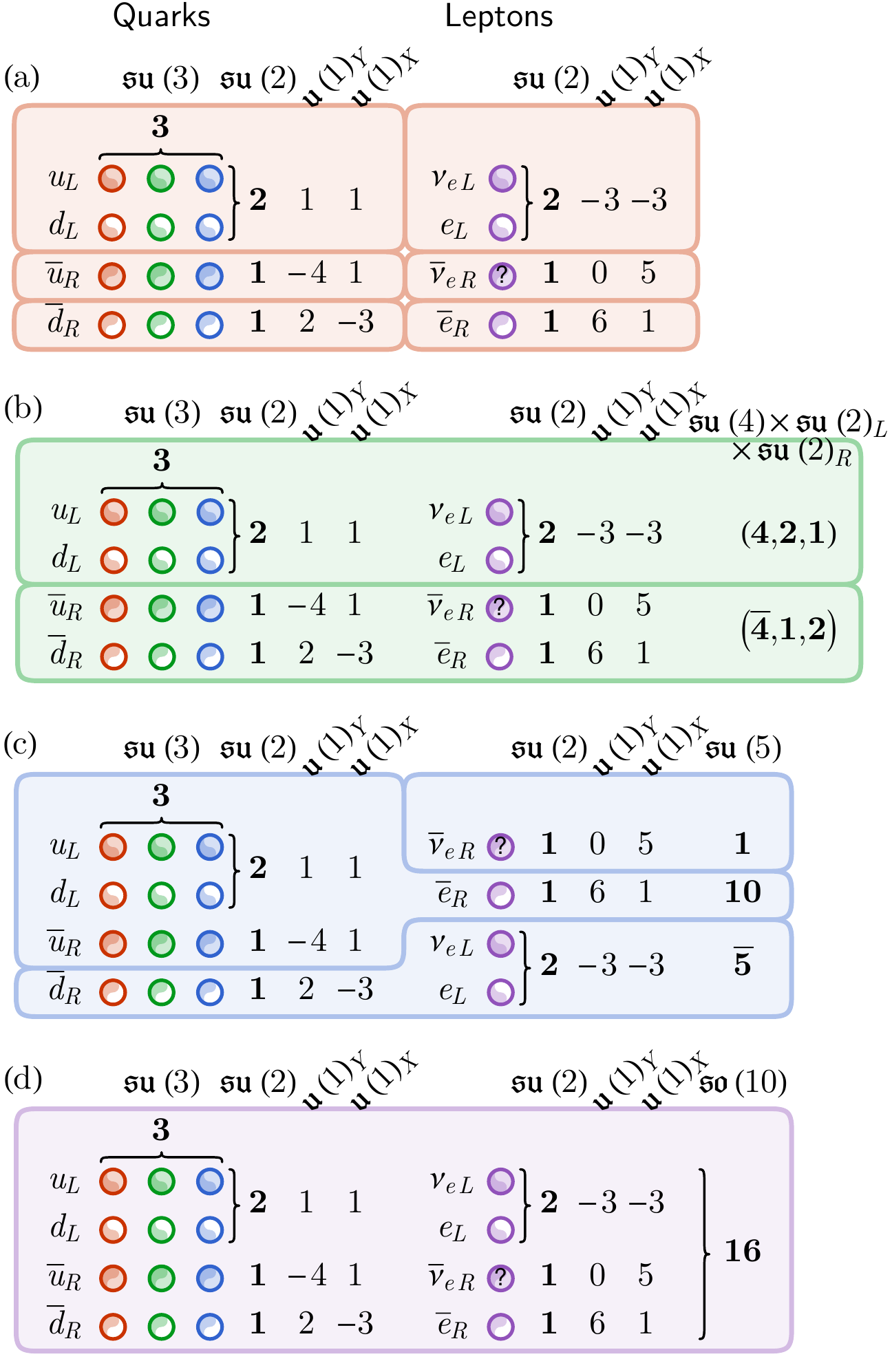}
\caption{The first generation of matter particles in (a) the Standard Model (SM), (b) the Pati-Salam (PS) model, i.e.~the $\so(6)\oplus\so(4)$ GUT, (c) the Georgi-Glashow (GG) model, i.e.~the $\u(5)$ GUT, and (d) the $\so(10)$ GUT. $Y$ denotes the electroweak hypercharge, and $X=5(B-L)-\frac{2}{3}Y$
with $(B-L)$ as the baryon minus lepton number.}
\label{fig:models}
\end{center}
\end{figure}

The Standard Model (SM) \cite{Glashow1961trPartialSymmetriesofWeakInteractions, Salam1964ryElectromagneticWeakInteractions, Salam1968, Weinberg1967tqSMAModelofLeptons} of particle physics summarizes our current knowledge of the fundamental building blocks of our universe in the 4-dimensional spacetime (4d). It describes three generations of fermionic matter particles interacting with gauge forces and an electroweak Higgs field. In each generation, if we assume that there exists a right-handed neutrino, there are sixteen Weyl fermion matter particles, containing left- and right-handed quarks (three colors and two flavors) and leptons (two flavors), as summarized in \figref{fig:models}(a). One of the open questions about the SM is whether the three fundamental gauge forces (electromagnetic, weak, and strong) are actually manifestations of a single all-encompassing force. This is the dream of the Grand Unified Theory (GUT). Throughout the history, many GUT models has been proposed, including the Pati-Salam (PS) model (or the $\so(6)\times\so(4)$ GUT) \cite{Pati1974yyPatiSalamLeptonNumberastheFourthColor} , the Georgi-Glashow (GG) model (or the $\su(5)$ GUT) \cite{Georgi1974syUnityofAllElementaryParticleForces} , and the $\so(10)$ GUT \cite{Fritzsch1974nnMinkowskiUnifiedInteractionsofLeptonsandHadrons}. The pattern that matter particles are unified under these GUTs are illustrated in \figref{fig:models}(b-d) respectively. 

\begin{figure}[!htbp]
\begin{center}
\includegraphics[width=\columnwidth]{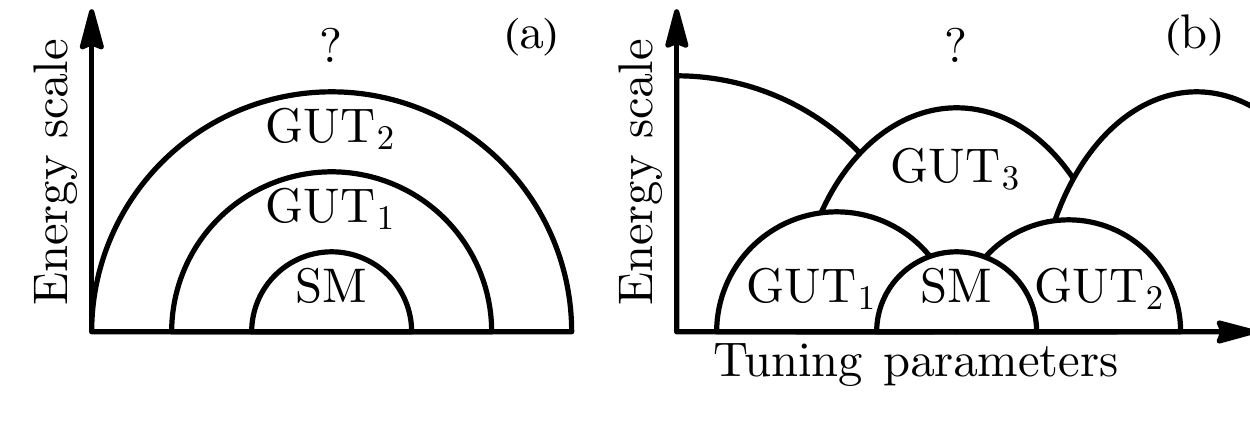}
\caption{Two different perspectives of how GUTs can be accessed from the SM: (a) high energy physics perspective, (b) condensed matter perspective. Experiments like proton decay only rules out the perspective (a), but not (b).}
\label{fig:perspectives}
\end{center}
\end{figure}

Despite the theoretical elegance of GUTs, experiments (such as proton decay \cite{Nath2007Proton}) have ruled out many (non-supersymmetric) GUTs as the high-energy completion of the SM, meaning that these GUTs might not be achieved only by raising up to a higher energy scale (or higher temperature scale) from the SM through thermal phase transitions, as shown in \figref{fig:perspectives}(a). However, experiments did not rule out the possibilities that these GUTs can be accessed by tuning model parameters away from the SM through quantum phase transitions, as shown in \figref{fig:perspectives}(b). From the high energy physics perspective, different GUTs are just different quantum field theories describing how gauge forces can be unified at higher energy scales. However, in condensed matter physics, different quantum field theories are often considered low-energy effective descriptions of different quantum phases of matter. From this perspective, the SM and various GUT vacua may be viewed as competing quantum phases of our universe, such that it becomes natural to discuss quantum phase transitions (and the associated quantum criticality) among different GUT vacua. By quantum criticality, we mean to emphasize the near zero-temperature 
quantum physics of gapless, scale or conformal invariant critical phenomena \cite{subirsachdev2011book}.
The quantum vacua (phases) tuning parameters appear perturbatively irrelevant at the SM fixed point in the renormalization group (RG) setup, hence their non-perturbative effects are largely overlooked in the previous literature. Exploring the effects of these tuning parameters and investigating quantum phase transitions among the SM and GUTs has motivated a series of works recently \cite{Wang2021Gauge1,Wang2021Gauge2,Wang2021Cobordism}.

The major differences between these field theories are their gauge groups. Therefore, the phase transition between GUTs (and the SM) are generally gauge-Higgs transitions, which is driven by the condensation of certain GUT Higgs field (distinct from the electroweak Higgs field in the SM) to Higgs down the gauge group. (An important remark: The present work only focuses on the GUT Higgs fields and their induced phase transitions; we do not consider the electroweak Higgs field or
electroweak symmetry breaking below the electroweak energy scale within SM.
So whenever Higgs fields are mentioned, 
they stand for the GUT Higgs fields for GUTs
instead of the electroweak Higgs field within the SM.)
If one conceptually suppresses the gauge fluctuation and treats the gauge group $G$ to be a global internal symmetry in $G$, the gauge-Higgs transitions will be closely analogous to spontaneous symmetry breaking transitions. {Therefore below we will switch between the gauge group $G$ and the global internal symmetry group $G$ perspectives,
based on the Weyl's principle of ``gauging'' or ''ungauging'' via ``summing over'' or ``fixing'' the $G$-bundle with $G$-connections in the field theory path integral,
depending on the context. Besides, in this article, we intentionally 
simplify the discussion
of the global symmetry by focusing only on the internal symmetry $G$
but omitting the spacetime symmetry (see the full 
spacetime-internal symmetry in \cite{Wang2021Gauge1,Wang2021Gauge2,Wang2021Cobordism}).}

The study of symmetry-breaking transitions has a long history in condensed matter physics \cite{Anderson1972More}. The conventional symmetry breaking phases and their phase transitions are successfully described by the Landau-Ginsburg-Wilson (LGW) paradigm in terms of order parameters \cite{Landau1958}. 
However, the condensed matter literature has also seen much discussion \cite{SenthildQCP0311326,Senthil2004Quantum,Motrunich2004Emergent,LevinSenthil0405702,Sandvik2007Evidence,Sachdev2008Quantum,Sachdev2009Exotic,Vishwanath2013Physics} on exotic quantum phase transitions beyond the LGW paradigm, 
{one notable example is} the deconfined quantum criticality (DQC) \cite{SenthildQCP0311326,Senthil2004Quantum,Motrunich2004Emergent}. The DQC occurs in a quantum system with a global symmetry $G$ that has two spontaneous symmetry breaking phases which break $G$ to its distinct subgroups $G_1$ and $G_2$ respectively. When the low-energy effective theory has a 't Hooft anomaly of $G$, the two symmetry breaking phases can not share a trivially gapped $G$-symmetric intermediate phase, resulting in a gapless critical point {(or a gapless critical region as an intermediate new phase)} between the two symmetry breaking phases. The theory for the quantum critical point/phase is usually not expressed in terms of order parameters in $G/G_1$ or $G/G_2$, but in terms of ``fractionalized'' degrees of freedom (such that the order parameters are composite of deconfined fractionalized particles, i.e.~partons), and hence the terminology ``deconfined''.

This article reviews the recent progress \cite{Wang2021Gauge1} in exploring the possibility of DQC in quantum phase transitions between different GUT vacua. The study reveals the SM as a neighboring phase of two competing GUT phases (i.e.~PS and GG). The competition could give rise to an intermediate DQC phase where the GUT Higgs field fractionalizes to deconfined particles and emergent gauge fields beyond the SM. Some of the fractionalized particles may persist in the SM phase, which contributes to dark matter candidates of our universe.

\section{GUT Higgs Condensation Transitions among GUT Phases}

The Yang-Mills gauge theory \cite{Yang1954Conservation}, or more generally the non-Abelian gauge theory, provides the basis for the theoretical formulation of the SM and GUTs. In this framework, the {matter} particles are excitations of Weyl fermion fields, and the gauge forces are mediated by non-Abelian gauge fields. The action of the Yang-Mills theory {coupled to Weyl fermions}
in the 4d spacetime generally takes the form of
\eq{\label{eq:S_YM}S_\text{YM}=\int_{\scM^4} \big(\dd^4 x \; \psi^\dagger(\ii\bar{\sigma}^\mu D_{\mu,A})\psi\big)
+\Tr(F\wedge \star F) ,}
where $\psi$ denotes the Lorentz spinor Weyl fermion field, the covariant derivative {$D_{\mu,A}=\partial_\mu-\ii g A_\mu$ contains the Lie algebra-valued gauge field $A_\mu$, and $F=\dd A-\ii g A\wedge A$ is the gauge curvature 2-form (given $A=A_\mu\dd x^\mu$) with the Yang-Mills gauge coupling $g$}. All Weyl fermions in the theory can be taken to have the 
left-handed helicity, with the choice of $\bar{\sigma}^\mu=(\sigma^0,-\sigma^1,-\sigma^2,-\sigma^3)$, as right-handed fermions can always be conjugated to their left-handed antiparticles. Each generation of matter particles consists of sixteen or fifteen Weyl fermions. In particular, the 16th Weyl fermion is the right-handed neutrino \cite{Drewes2013The-Phenomenology,Boyarsky2019Sterile} which is sterile to the SM gauge force and has not been observed in experiments so far. However, there is a good reason to include the sterile neutrino, as it provides the simplest way to cancel a $\dsZ_{16}$ global anomaly of the theory \cite{Wen2013ppa1305.1045,You2015Interacting,BenTov2015graZee1505.04312,
GarciaEtxebarriaMontero2018ajm1808.00009, WW2019fxh1910.14668}. Otherwise, with only fifteen Weyl fermions, the anomaly must be canceled in more exotic ways \cite{Wang2020Anomaly, JW2008.06499, JW2012.15860}, either by additional 4d or 5d gapped topological quantum field theories (TQFTs) or by 4d gapless interacting conformal field theories (CFTs).

Assuming sixteen Weyl fermions in each generation, which couple to gauge fields that mediate gauge interactions, the main difference among different GUT models relies on their different choices of the gauge group $G$ (or the gauge algebra $\g$), summarized as follows:
\begin{itemize}
\item SM: $G_\text{SM}=(\SU(3)\times\SU(2)_L)\times_{\dsZ_6}\U(1)_Y\times_{\dsZ_6}\U(1)_X$ (where $\SU(3)$ denotes the color group, $\SU(2)_L$ denotes the weak isospin group, $\U(1)_Y$ denotes the electroweak hypercharge group, and {$\U(1)_X$ denotes the $X  \equiv 5(B-L)-\frac{2}{3}Y$-charge \cite{Wilczek1979hcZee} group
which is a variant of baryon minus lepton $(B-L)$-like symmetry}), correspondingly, $\g_\text{SM}=\su(3)\oplus\su(2)_L\oplus\u(1)_Y\oplus\u(1)_X$. {Note that
we introduce this extra  
$s\u(1)_X$ into $\g_\text{SM}$.
} 

\item PS: $G_\text{PS}=\Spin(6)\times_{\dsZ_2}\Spin(4)$, correspondingly, $\g_\text{PS}=\so(6)\oplus\so(4)=\su(4)\oplus\su(2)_L\oplus\su(2)_R$.

\item GG: $G_\text{GG}=\SU(5)\times_{\dsZ_{5|2}}\U(1)_{X}=\U(5)_{2}$ (where the subscript $2$ indicates that the $\dsZ_5$ center of $\SU(5)$ is identified with the $\dsZ_5$ subgroup of $\U(1)_X$ by a double covering, or more precisely defined by the short exact sequence $1\to\SU(5)\to\U(5)_m\xrightarrow{\det}\U(1)/\dsZ_{5m}\to 1$ \cite{Wang2021Gauge2}), correspondingly, $\g_\text{GG}=\su(5)\oplus\u(1)_X$.

\item $\so(10)$ GUT: $G_{\so(10)}=\Spin(10)$, correspondingly, $\g_{\so(10)}=\so(10)$.
\end{itemize}
The notation $G_1\times_{N}G_2\equiv\frac{G_1\times G_2}{N}$ denotes the product of two groups $G_1$ and $G_2$ quotient their common normal subgroup $N$. One may as well treat these groups as global symmetry groups by ``un-gauging'' the gauge theory, i.e.~by considering the dynamical gauge field as a static background probe. This allows us to understand the transitions between these models from the perspective of spontaneous symmetry breaking. 

\begin{figure}[htbp]
\begin{center}
\includegraphics[width=0.52\columnwidth]{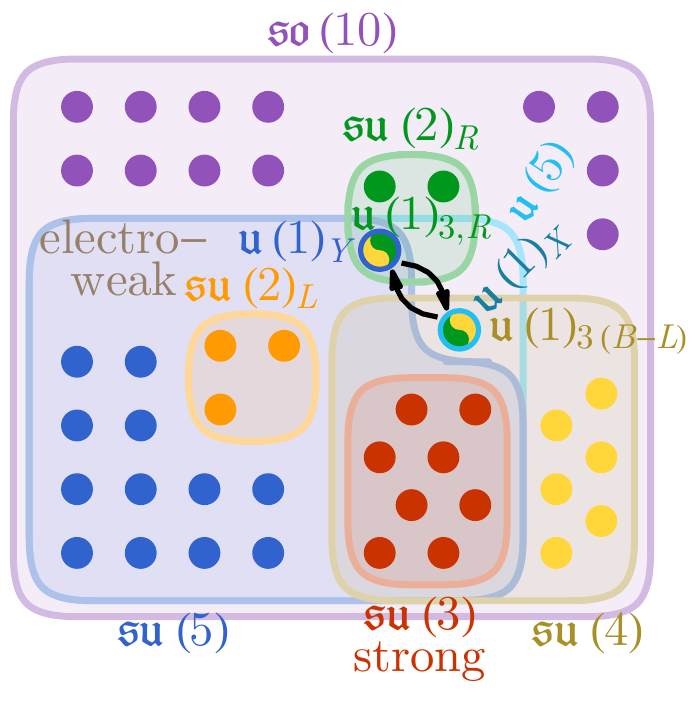}
\caption{Unification of gauge forces as Lie algebra embedding. Each dot represents a gauge boson (corresponding to a Lie algebra generator). $\u(1)_Y\oplus\u(1)_X$ and $\u(1)_{3,R}\oplus\u(1)_{3(B-L)}$ generators are linearly mixed to accommodate different unification schemes, see \eqnref{eq:u1mixing}.}
\label{fig:algebra}
\end{center}
\end{figure}

A phase with a larger symmetry group $G_\text{large}$ can undergo a symmetry breaking transition to a phase with a smaller symmetry group $G_\text{small}$, if $G_\text{small}$ is a subgroup of $G_\text{large}$, or equivalently if $G_\text{small}$ can be embedded in $G_\text{large}$ (denoted by $G_\text{small}\hookrightarrow G_\text{large}$). The group of the SM and various GUTs can be embedded as \cite{BaezHuerta0904.1556}
\eq{\label{eq:embed}\begin{matrix}
G_\text{GG} & \hookrightarrow & G_{\so(10)} \\
\hookuparrow & & \hookuparrow\\
G_\text{SM} & \hookrightarrow & G_\text{PS}
\end{matrix}.}
The embedding structure indicates that the $\so(10)$ GUT can go through symmetry-breaking transitions from
the $\so(10)$ GUT phase to the adjacent PS or GG phases. The PS and GG phases can further break symmetry to their common adjacent SM phase. The Lie group embedding structure in \eqnref{eq:embed} also applies to their corresponding Lie algebras, as illustrated in \figref{fig:algebra}. It turns out that $\g_\text{SM}$ is the unique common subalgebra of $\g_\text{PS}$ and $\g_\text{GG}$, while $\g_{\so(10)}$ is the minimal algebra that contains both $\g_\text{PS}$ and $\g_\text{GG}$:
\eqs{\g_\text{PS}\cap\g_\text{GG}&=\g_\text{SM},\\
\g_\text{PS}\cup\g_\text{GG}&=\g_{\so(10)}.\\}
{More than just Lie algebra intersection/union,
there are also corresponding statements in Lie groups
\cite{BaezHuerta0904.1556,Wang2021Gauge1,Wang2021Gauge2}.}
The stringent relations between these symmetry groups (or algebras) suggest that there might be a unifying theory that describes the phase transitions among the SM and different GUTs.

It is worth mentioning that the subalgebra $\u(1)_Y\oplus\u(1)_X\cong \u(1)_{3,R}\oplus\u(1)_{3(B-L)}$ inside $\g_\text{SM}$ is shared between $\g_\text{PS}$ and $\g_\text{GG}$, but it can have different choices of generator basis. Let $Y\in\dsZ$ be the weak hypercharge (at triple-scale), $X\in\dsZ$ be the $X$-charge, $Q_{3,R}=2I_{3,R}\in\dsZ$ be the right-isospin (at double-scale), $(B-L)\in\dsZ/3$ be the baryon minus lepton number. They are related by the linear relation
\eq{\label{eq:u1mixing} \mat{Y\\X}=\mat{3&1\\-2&1}\mat{Q_{3,R}\\3(B-L)}.}
The electromagnetic $\u(1)_\text{em}$ charge is given by $Q_\text{em}=I_{3,L}+\frac{1}{6}Y=I_{3,L}+I_{3,R}+\frac{1}{2}(B-L)$, where $I_{3,L}\in\dsZ/2$ is the left-isospin (the weak isospin).

Within the Landau-Ginzburg-Wilson paradigm, the symmetry-breaking transition from a larger symmetry group $G_\text{large}$ (or its algebra $\g_\text{large}$) to a smaller symmetry group $G_\text{small}$ (or its algebra $\g_\text{small}$) can be formulated as the condensation of the order parameter field, whose target manifold is $G_\text{large}/G_\text{small}$. If the symmetries are dynamically gauged, the symmetry-breaking transition can be reinterpreted as the Higgs transition, in which the order parameter field is promoted to the Higgs field. The Higgs field that drives the transition between GUT (and SM) phases will be called the GUT Higgs fields to distinguish from the electroweak Higgs field in the SM. In order for the Higgs condensation to break $\g_\text{large}$ but not $\g_\text{small}$, it must be in a non-trivial representation of $\g_\text{large}$ but in the trivial representation of $\g_\text{small}$. One can look up such cases from Lie algebra branching rules. 

For example, to break $\g_{\so(10)}=\so(10)$ to $\g_\text{PS}=\su(4)\oplus\su(2)_L\oplus\su(2)_R$, the branching rule $\mathbf{54}\to(\mathbf{1},\mathbf{1},\mathbf{1})+\cdots$ suggests that one could condense a GUT Higgs field $\Phi_\mathbf{54}$ in the $\mathbf{54}$ representation of $\g_{\so(10)}$ to the $(\mathbf{1},\mathbf{1},\mathbf{1})$ representation of $\g_\text{PS}$ to achieve the desired symmetry breaking. Similarly, to break $\g_{\so(10)}$ to $\g_\text{GG}=\su(5)\oplus\u(1)_X$, the branching rule $\mathbf{45}\to\mathbf{1}_{0}+\cdots$ suggests that one could condense a GUT Higgs field $\Phi_\mathbf{45}$ in the $\mathbf{45}$ representation of $\g_{\so(10)}$ to the $\mathbf{1}_{0}$ representation of $\g_\text{GG}$ to achieve the desired symmetry breaking. Furthermore, if both GUT Higgs fields $\Phi_\mathbf{45}$ and $\Phi_\mathbf{54}$ are condensed, the internal symmetry will be broken down to $\g_\text{SM}$, as $\g_\text{SM}$ is the intersection of $\g_\text{PS}$ and $\g_\text{GG}$ that survives both symmetry breaking. More concretely, starting from the PS phase where the GUT Higgs field $\Phi_\mathbf{54}$ is already condensed, to further break $\g_\text{PS}$ to $\g_\text{SM}=\su(3)\oplus\su(2)_L\oplus\u(1)_Y\oplus\u(1)_X$, the branching rules $(\mathbf{1},\mathbf{1},\mathbf{3})\to(\mathbf{1},\mathbf{1})_{0,0}+\cdots$ and $(\mathbf{15},\mathbf{1},\mathbf{1})\to(\mathbf{1},\mathbf{1})_{0,0}+\cdots$ in the $\mathbf{45}$ sector indicates that the GUT Higgs field $\Phi_\mathbf{45}$ needs to be condensed as well (more precisely, a certain combination of {$(\mathbf{1},\mathbf{1},\mathbf{3})$ and $(\mathbf{15},\mathbf{1},\mathbf{1})$ should be condensed, whose specific form will be discussed later).} On the other hand, starting from the GG phase where the GUT Higgs field $\Phi_\mathbf{45}$ is already condensed, to further break $G_\text{GG}$ to $G_\text{SM}$, the branching rule $\mathbf{24}_{0}\to(\mathbf{1},\mathbf{1})_{0,0}+\cdots$ in the $\mathbf{54}$ sector indicates that the GUT Higgs field $\Phi_\mathbf{54}$ needs to be condensed as well. Both analyses points to the same conclusion that the SM phase can be achieved from the $\so(10)$ GUT phase by simultaneously condensing the two GUT Higgs fields $\Phi_\mathbf{45}$ and $\Phi_\mathbf{54}$. 

\begin{figure}[htbp]
\begin{center}
\includegraphics[width=0.55\columnwidth]{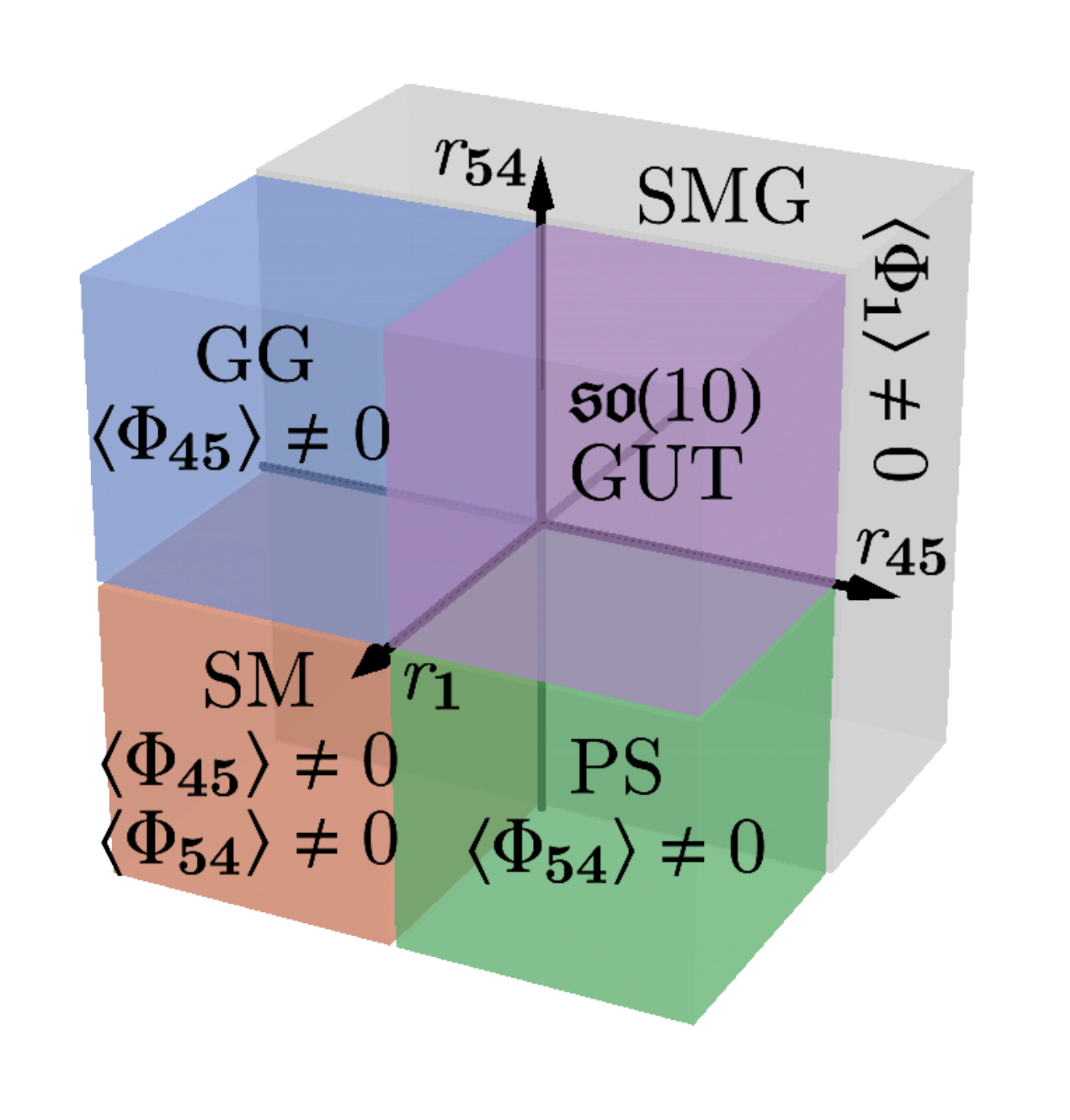}
\caption{Schematic quantum phase diagram of GUT phases adjacent to the SM phase. Phase labels stand for: SM - the Standard Model phase, PS - the Parti-Salam $\so(6)\oplus\so(4)$ phase, GG - the Georgi-Glashow $\u(5)$ phase, and SMG - the symmetric mass generation phase.}
\label{fig:phase}
\end{center}
\end{figure}

To connect different GUT phases, one can start with the largest internal symmetry (or gauge) group $G_{\so(10)} = \Spin(10)$ and access lower-symmetry phases by Higgs condensation. The Higgs condensation can be described by 
\eq{\label{eq:S_H}S_\text{H}=\int_{\scM^4}\dd^4x
\big(\sum_{\mathbf{R}}(D_{\mu,A}\Phi_\mathbf{R})^2+V(\Phi_\mathbf{R})\big),}
where $\Phi_\mathbf{R}$ is the Lorentz scalar GUT Higgs field in the $\mathbf{R}$ representation of $\so(10)$, and the Higgs potential takes the form of $V(\Phi_\mathbf{R})=r_\mathbf{R}\Phi_\mathbf{R}^2+\lambda_\mathbf{R}\Phi_\mathbf{R}^4+\cdots$. The $\Phi^4$ coefficient $\lambda_\mathbf{R}>0$ is always assumed to be positive. The $\Phi^2$ coefficients $r_\mathbf{R}$ is the tuning parameter that drives the transition between GUT phases. At the mean-field level, when $r_\mathbf{R}<0$, the corresponding Higgs field $\Phi_\mathbf{R}$ will condense, i.e.~$\langle\Phi_\mathbf{R}\rangle\neq0$. Based on the above discussion, a schematic phase diagram can be concluded as in \figref{fig:phase} (on the $r_\mathbf{1}>0$ side).

The GUT Higgs field in both the $\mathbf{45}$ and $\mathbf{54}$ representation motivates us to include an additional GUT Higgs field $\Phi_\mathbf{1}$ in the $\mathbf{1}$ (trivial) representation, such that $\mathbf{1}+\mathbf{45}+\mathbf{54}$ can be unified as the $\so(10)$ bivector representation $\mathbf{10}\times\mathbf{10}$, while the $\so(10)$ vector representation $\mathbf{10}$ can be further composed by the product of spinor representations as $\mathbf{16}\times\mathbf{16}\to\mathbf{10}+\mathbf{120}+\overline{\mathbf{126}}$. This enables us to write down the most natural Yukawa-like coupling between the GUT Higgs field and the matter fermion field (in the $\mathbf{16}$ representation),
\eqs{\label{eq:S_Y}S_\text{Y}&=\int_{\scM^4}\dd^4 x\frac{1}{2}\Big(\phi^\intercal\Phi\phi\\
&+\sum_{a=1}^{5}\big(\psi^\intercal\ii\sigma^2(\phi_{2a-1}\Gamma_{2a-1}-\ii\phi_{2a}\Gamma_{2a})\psi+\text{h.c.}\big)\Big).}
Here the GUT Higgs field $\Phi$ is written in the $\so(10)$ bivector representation as a $10\times 10$ matrix field (with totally 100 real scalar components). The $\so(10)$ vector field $\phi$ has 10 real scalar components, which corresponds to the 10 orthogonal Majorana masses (pairing terms) among 16 Weyl fermions. $\Gamma_a$ (for $a=1,2,\cdots,10$) are ten rank-16 real symmetric matrices satisfying $\{\Gamma_{2a-1},\Gamma_{2b-1}\}=2\delta_{ab}$, $\{\Gamma_{2a},\Gamma_{2b}\}=2\delta_{ab}$, $\{\Gamma_{2a-1},\Gamma_{2b}\}=0$ (for $a,b=1,2,\cdots,5$). In \eqnref{eq:S_Y}, $\ii\sigma^2$ acts in the 2-dimensional spacetime spinor subspace and $\Gamma_a$ acts in the 16-dimensional flavor subspace. Unlike the electroweak Higgs field in the SM, which gives quadratic mass to the matter fermions, the GUT Higgs field $\Phi$ does not generate fermion-bilinear mass, but rather serves as a four-fermion interaction. This can be seen by treating the $\so(10)$ vector field $\phi$ as a Lagrangian multiplier. Integrating out $\phi$ will leads to the quartic coupling of the form $\Phi \psi\psi \psi\psi$. As four-fermion interactions are perturbatively irrelevant for 4d Weyl fermions, perturbative vacuum expectation value of the GUT Higgs field (i.e.~$\langle\Phi\rangle\neq 0$ but small) will not affect the matter fermion spectrum. 

As an $\so(10)$ bivector representation $\mathbf{10}\times\mathbf{10}$, the GUT Higgs field $\Phi$ can be decomposed into $\mathbf{1}+\mathbf{45}+\mathbf{54}$ sectors, which defines
\eqs{\Phi_\mathbf{1}&: \Tr\Phi=\sum_{a}\Phi_{aa},\\
\Phi_\mathbf{45}&: \Phi_{[a,b]}=\tfrac{1}{2}(\Phi_{ab}-\Phi_{ba}),\\
\Phi_\mathbf{54}&: \Phi_{\{a,b\}}=\tfrac{1}{2}(\Phi_{ab}+\Phi_{ba}).\\}
Starting from the $\so(10)$ GUT, the condensation of $\Phi_\mathbf{45}$ or/and $\Phi_\mathbf{54}$ will drive symmetry breaking transitions to various lower-symmetry phases, as summarized in \figref{fig:phase}. In particular, if $\Phi$ condenses to a specific configuration $\langle\Phi\rangle$, the original algebra $\g_\text{large}$ will be broken to its subalgebra $\g_\text{small}=\{x\in\g_\text{large}|[x,\langle\Phi\rangle]=0\}$. It can be verified that the following condensation configurations
\eqs{\label{eq:Phi_config}\langle\Phi_\mathbf{54}\rangle&\propto\smat{-3&&&&\\&-3&&&\\&&2&&\\&&&2&\\&&&&2}\otimes\smat{1&\\&1},\\
\langle\Phi_\mathbf{45}\rangle&\propto \smat{1&&&&\\&1&&&\\&&1&&\\&&&1&\\&&&&1}\otimes\smat{0&1\\-1&0},}
induce the following symmetry breaking
\begin{equation}
\begin{tikzcd}
                                & \g_{so(10)} \arrow[ld, "\langle\Phi_\mathbf{54}\rangle\neq0"'] \arrow[rd, "\langle\Phi_\mathbf{45}\rangle\neq0"] &                                \\
\g_\text{PS} \arrow[rd, "\langle\Phi_\mathbf{45}\rangle\neq0"'] &                                                         & \g_\text{GG} \arrow[ld, "\langle\Phi_\mathbf{54}\rangle\neq0"] \\
                                & \g_\text{SM}                                                   &                                
\end{tikzcd}
\end{equation}
$\langle\Phi_\mathbf{54}\rangle$ explicitly distinguishes the first four-dimensional subspace from the last six-dimensional subspace of the $\so(10)$ vector, therefore breaking $\g_{\so(10)}=\so(10)$ down to $\g_\text{PS}=\so(6)\oplus\so(4)$. $\langle\Phi_\mathbf{45}\rangle$ is proportional to the $\u(1)_X$ generator, which effectively requires the unbroken generators to commute with $\u(1)_X$ generator, which singles out $\g_\text{GG}=\su(5)\oplus\u(1)_X$. When both of them condense, the symmetry is broken to $\g_\text{PS}\cap\g_\text{GG}=\g_\text{SM}$.

The condensation of $\Phi_\mathbf{1}$ in the trivial sector will not break the $G_{\so(10)}$ symmetry, and a perturbative vacuum expectation value of $\langle \Phi_\mathbf{1}\rangle < \Phi_{\mathbf{1},c}$ {below a critical value} also has no effect on the fermion spectrum, so the theory remains in the $\so(10)$ GUT phase. However, when the condensation $\langle \Phi_\mathbf{1}\rangle > \Phi_{\mathbf{1},c}$ exceeds a critical value, the quartic interaction induced by $\langle \Phi_\mathbf{1}\rangle$ can have  non-perturbative effects: it can generate an interacting gap for all fermions without breaking the $G_{\so(10)}$ symmetry. The mechanism is known as the symmetric mass generation (SMG), and has been explored in various situations, as some selective examples, for fermion zero modes in 1d \cite{FidkowskifSPT2}, for chiral fermions in 2d \cite{Wang2013ytaJW1307.7480,Wang2018ugfJW1807.05998}, for Dirac fermions in 3d \cite{YouHeXuVishwanath1705.09313, YouHeVishwanathXu1711.00863}, and most notable for Weyl fermions in 4d \cite{Eichten1985ftPreskill1986,Wen2013ppa1305.1045,You2015Interacting,BenTov2015graZee1505.04312,Kikukawa2017ngf1710.11618,Catterall2020fep, CatterallTogaButt2101.01026,RazamatTong2009.05037, Tong2104.03997}. To consider the possibility of SMG, one can investigate the strongly interacting limit of $\langle \Phi_\mathbf{1}\rangle \to\infty$, where the {matter} fermion dynamics is dominated by the interaction $S_\text{Y}$ in \eqnref{eq:S_Y}. In this limit, the field theory decouples in each spatial location and can be solved exactly. The solution shows that the many-body ground state at each location is unique with a finite excitation gap \cite{You2015Interacting}, indicating all fermions are gapped by the interaction without spontaneous symmetry breaking. 

\section{Deconfined Quantum Criticality}

The phase diagram in \figref{fig:phase} resembles the phase diagram of deconfined quantum criticality in (2+1)D quantum magnet \cite{SenthildQCP0311326,Senthil2004Quantum,Vishwanath2013Physics}, where two distinct symmetry breaking phases -- the antiferromagnetic N\'eel phase and the valence bond solid (VBS) phase -- are connected by a direct continuous quantum phase transition \cite{Sandvik2007Evidence}. The N\'eel phase spontaneously breaks the spin $\SO(3)$ rotation symmetry and the VBS phase spontaneously breaks the lattice rotation $\mathbb{Z}_4$ symmetry (or $\SO(2)$ symmetry in the continuum). The direct transition between these two quantum phases is exotic, which involves fractionalized excitations and emergent gauge fields. The deconfinement of the fractionalized excitations happens at and only at the N\'eel-VBS critical point, therefore the critical point is also called the deconfined criterial point (DQCP). In the phase diagram \figref{fig:phase}, the PS and GG phases also breaks the $G_{\so(10)}=\Spin(10)$ symmetry differently, which motivates the investigation of the possible DQCP physics along the phase boundary between the PS and GG phases.

The key mechanism of DQCP is that the topological defect of one order parameter carries the charge of the other order parameter. In this case, the two order parameters are said to be intertwined orders. For example, the vortex of VBS order parameter carries a spin-1/2 degree of freedom that is charged under the spin $\SO(3)$ symmetry in
a fractionalized and projective representation \cite{LevinSenthil0405702}. 
To explore similar scenarios in the GUTs, one should first understand the possible topological defects of the order parameters in the PS and GG phases. When the GUT Higgs fields $\Phi_\mathbf{45}$ and $\Phi_\mathbf{54}$ condense to a finite amplitude, their orientational fluctuations belongs to the following target manifolds
\eq{\Phi_\mathbf{45}\in\frac{\SO(10)}{\U(5)}, \quad \Phi_\mathbf{54}\in\frac{\SO(10)}{\SO(6)\times\SO(4)}.}
These manifolds have non-trivial second homotopy groups
\eq{\pi_2\Big(\frac{\SO(10)}{\U(5)}\Big)=\dsZ,\quad
\pi_2\Big(\frac{\SO(10)}{\SO(6)\times\SO(4)}\Big)=\dsZ_2,}
which indicates that the $\Phi_\mathbf{45}$ order parameter admits $\dsZ$-classified topological point defects in the 3D space, and the $\Phi_\mathbf{54}$ order parameter admits $\dsZ_2$-classified topological point defects in the 3D space. The intertwinement of these two symmetry-breaking orders is equivalently characterized by the non-trivial mutual braiding statistics of their topological defects, i.e.~the $\Phi_\mathbf{45}$ defect and the $\Phi_\mathbf{54}$ defect will see each other {as a monopole emitting totally $2\pi$ 
statistical Berry phase flux}.

\begin{figure}[htbp]
\begin{center}
\includegraphics[width=0.45\columnwidth]{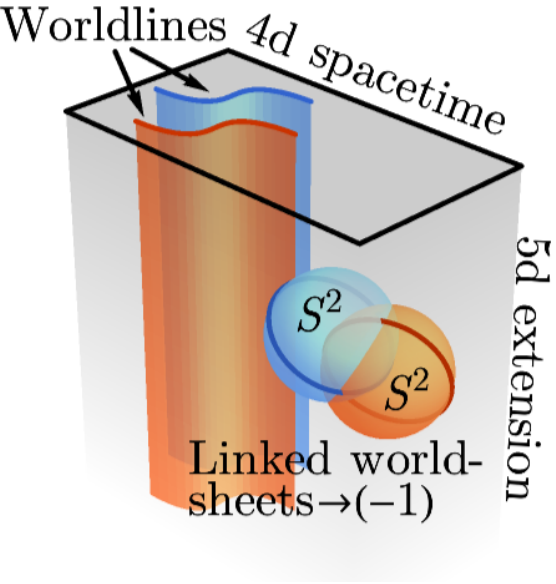}
\caption{Illustration of defect worldsheets in extended 5d spacetime. The two colors indicate the $\Phi_\mathbf{45}$ or the $\Phi_\mathbf{54}$ topological defects. Each pair of mutually linked 2-spheres of the two types of 
topological defects will contribute a $(-1)$ sign to the path integral weight.}
\label{fig:link}
\end{center}
\end{figure}

The braiding statistics of topological defects can be more precisely defined by extending the 4d spacetime to 5d with an auxiliary dimension. As shown in \figref{fig:link}, the topological defect world-lines in the 4d spacetime will extended to pseudo world-sheets in 5d. The defect braiding statistics corresponds to the effect that each pair of linked 2-spheres of the world-sheets of two different types of defects will contribute a $(-1)$ sign to the partition function
weight \cite{Putrov2016qdo1612.09298PWY,GuoJW1812.11959}. On the field theory level, this effect can be described by a Wess-Zumino-Witten (WZW) term, defined in the 5d extension, 
\eq{e^{\ii S_\text{WZW}[\Phi]}=\exp\Big(\ii\pi\int_{\scM_5}B(\Phi_\mathbf{54})\smile\delta B'(\Phi_\mathbf{45})\Big)\Big|_{\scM_4=\partial \scM_5},}
where $B$ and $B'$ are two-cocycles also two-cohomology class of the second cohomology group that evaluate the defect topological number when they are integrated over a 2-sphere enclosing the point defect in the 3D space,
\eqs{B&\in {\rm H}^2\Big(\frac{\SO(10)}{\SO(6)\times\SO(4)},\dsZ_2\Big)\\
&=\text{Hom}\Big(\pi_2\Big(\frac{\SO(10)}{\SO(6)\times\SO(4)}\Big),\dsZ_2\Big),\\
B'&\in {\rm H}^2\Big(\frac{\SO(10)}{\U(5)},\dsZ\Big)=\text{Hom}\Big(\pi_2\Big(\frac{\SO(10)}{\U(5)}\Big),\dsZ\Big).\\
}
{Here we use the fact that
if $K$ is a $(n-1)$-connected topological space, then 
${\rm H}^n (K, {\rm A}) = \text{Hom} ( \pi_d (K), {\rm A})$
for some abelian ${\rm A}$, by the 
Hurewicz theorem and universal coefficient theorem.}

With the additional WZW term, the field theory (which has the {spacetime-internal symmetry} $\Spin\times_{\dsZ_2}\Spin(10)$ structure) takes the form of
\eq{S=S_\text{YM}[\psi,A]+S_\text{H}[\Phi, A]+S_\text{Y}[\psi,\Phi]+\ii S_\text{WZW}[\Phi].}
This field theory has a $w_2w_3$ $\dsZ_2$-class mixed gauge-gravitational global anomaly \cite{WangWen2018cai1809.11171, WanWang2018bns1812.11967,WangWenWitten2018qoy1810.00844} captured by a 5d bulk invertible topological quantum field theory (invertible TQFT) 
\eqs{&\exp\Big(\ii\pi\int_{\scM_5}w_2(V_{\SO(10)})w_3(V_{\SO(10)})\Big)\\
=&\exp\Big(\ii\pi\int_{\scM_5}w_2(TM)w_3(TM)\Big),}
where $w_2,w_3$ denotes the Stiefel-Whitney classes of either the 
associated $V_{\SO(10)}$ vector bundle or the $TM$ tangent bundle. This anomaly originated purely from the GUT Higgs sector $\Phi$. The matter fermions $\psi$ does not contribute to it. In the presence of this anomaly, the GUT Higgs field should be viewed as the 4d boundary modes of a 5d invertible TQFT. Due to the anomaly, the $\Spin(10)$ symmetry can not be dynamically gauged on the 4d boundary alone. The $\Spin(10)$ gauge field must propagates into the 5d bulk.

The $w_2w_3$ anomaly only exists in the presence of the $G_{\so(10)}=\Spin(10)$ symmetry. It is lifted when the $\Spin(10)$ symmetry is broken to any subgroups of interest: $G_\text{PS}$, $G_\text{GG}$ or $G_\text{SM}$. Thus condensing either $\Phi_\mathbf{45}$ or $\Phi_\mathbf{54}$ GUT Higgs fields will break the $\Spin(10)$ symmetry and cancels the $w_2w_3$ anomaly. The resulting PS, GG, and SM phases are just ordinary spontaneous symmetry breaking phases (or Higgs phases if symmetries are gauged). What becomes non-trivial is the $\Spin(10)$ symmetric $\so(10)$ GUT phase. Due to the anomaly introduced by the WZW term, the GUT Higgs fields can not be trivially disordered (namely, trivially
and featurelessly gapped without ground state degeneracy associated with topological order or symmetry breaking order). One possibility is that the GUT Higgs field enters a ``spin liquid'' fractionalized
phase when the $\Spin(10)$ symmetry is restored, which happens either at the PS-GG transition point or in the $\so(10)$ GUT phase. 

\begin{figure}[htbp]
\begin{center}
\includegraphics[width=0.55\columnwidth]{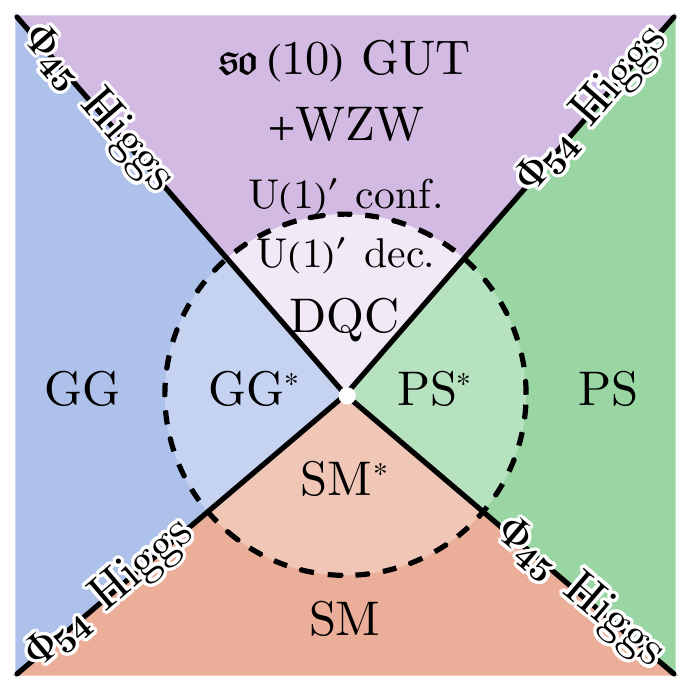}
\caption{Phase diagram in the presence of the WZW term. The dashed circle denotes the confine-deconfine transition of the emergent $\U(1)'$ gauge field.
The {solid-line} phase boundaries between two neighbor phases all are described by 
{Higgs-condensation 
continuous quantum phase} transitions.}
\label{fig:DQC}
\end{center}
\end{figure}

The fractionalized phase of the GUT Higgs field can be understood from a parton construction for the WZW field theory. The WZW term can be replaced by quantum electrodynamics (QED) theory in 4d spacetime, in terms of fractionalized fermionic partons $\xi$ coupled to an emergent $\U(1)'$ gauge field (this $\U(1)'$ is beyond the SM, 
not any of the $\U(1)$ gauge group in the SM or GUTs),
\eq{\label{eq:parton} S_{\text{QED}'_4}=\int_{\scM^4}\dd^4 x\;\bar{\xi} (\ii\gamma^\mu D_{\mu,a}-\Phi_\mathbf{54}-\ii\gamma^5\Phi_\mathbf{45})\xi.}
The fermionic parton field $\xi$ contains 10 Dirac fermions forming the representation $\mathbf{10}$ of $\SO(10)$, which also carries 1 unit of an emergent $\U(1)'$ gauge charge. The gauge field $a$ is neural under $\SO(10)$. The emergent $\U(1)'$ gauge bosons can be called ``dark'' photons as they do not couple to any of the SM particles. The dark photons mediate the gauge interaction between fermionic partons and bind them back into the GUT Higgs field $\Phi$. The GUT Higgs field $\Phi$ is in the $\mathbf{10}\times\mathbf{10}$ bivector representation of $\SO(10)$, which is fractionalized as the fermionic parton bilinear mass terms
\eq{\Phi_\mathbf{54} \sim \bar{\xi}\xi, \quad \Phi_\mathbf{45}\sim\bar{\xi}\ii\gamma^5\xi.} 
The phase described by the QED$'_4$ parton theory is similar to the $\U(1)$ Dirac spin liquid \cite{Zhou2017Quantum,Savary2017Quantum} discussed in the condensed matter physics context.

To check that the proposed parton theory \eqnref{eq:parton} reproduces the same $w_2w_3$ anomaly as the WZW term, one can temporally {turn off the 
GUT Higgs coupling to partons}, which allows the gauge group to be enlarged to $\SU(2)'$ (with $\U(1)'$ being its subgroup). Now the theory $\int_{\scM^4}\dd^4 x\;\bar{\xi} (\ii\gamma^\mu D_{\mu,a})\xi$ contains $2\times 10=20$ Weyl fermions, transforming as $(\mathbf{2},\mathbf{10})$ in $\SU(2)'\times\SO(10)$. The anomaly can be matched by tracking the fermion representations
\eq{
\begin{array}{ccccc}
\U(1)'\times\SO(10) & \hookrightarrow
&\SU(2)'\times\SO(10) & \hookrightarrow
& \Sp(10)\\
{\bf 10}_1 & & ({\bf 2},{\bf 10}) & \sim & {\bf 20}\\
& & & & \rotatebox[origin=c]{-90}{$\cong$} \\
\SU(2)''\times\Sp(8) & \hookrightarrow & \Sp(2)\times\Sp(8) & \hookrightarrow & \Sp(10) \\
({\bf 4},{\bf 1}) \oplus ({\bf 1},{\bf 16}) & \sim & ({\bf 4},{\bf 1}) \oplus ({\bf 1},{\bf 16}) & \sim & {\bf 20}
\end{array}.
}
The twenty Weyl fermions can be further embedded in the $\mathbf{20}$-dim representation of $\Sp(10)$, which transforms the fermion fields as ten quaternion Grassmann numbers. On the other hand, it can be split to $\Sp(2)\times\Sp(8)$ that separately rotate the first two and the last eight quaternion components $({\bf 4},{\bf 1}) \oplus ({\bf 1},{\bf 16})$. The $\Sp(2)$ group has an $\SU(2)''$ subgroup as its maximal special embedding. It is known that odd number of fermions in the $\mathbf{4}$-dim (spin-3/2) representation of $\SU(2)$ can generate the new $\SU(2)$ anomaly \cite{WangWenWitten2018qoy1810.00844}, described by the $w_2w_3$ anomaly term. Using this fact, the $\SU(2)''$ sector will give rise to the $w_2w_3$ anomaly and the $\Sp(8)$ sector will not (as it contains an even number of spin-3/2 fermions when broken down to its $\SU(2)$ subgroup similarly). By anomaly matching, one can conclude that the fermions will generate the $w_2w_3$ anomaly of the $\Sp(10)$ bundle. However, the emergent {$\U(1)'$ sector
of fermionic partons alone} does not give rise to the $w_2w_3$ anomaly, so the anomaly must be contained in the {$\SO(10)$ sector of fermionic partons}, which reproduces the anomaly of the WZW term.

The anomaly matching argument indicates that the gapless fermionic partons can induce the desired braiding statistics between the topological defects of $\Phi_\mathbf{45}$ and $\Phi_\mathbf{54}$, which provides a possible state that saturates the anomaly. In this state, the Higgs field $\Phi$ deconfines into fermionic partons coupled by the emergent $\U(1)'$ gauge field governed by the QED$'_4$ theory in \eqnref{eq:parton}. Na\"ively and intuitively, the GG and PS phases can then be connected by a continuous transition with a deconfined quantum critical (DQC) point, as the white dot in \figref{fig:DQC}. However, since the deconfined $\U(1)'$ gauge theory is stable in 4d, the critical point will extend into a critical phase in the $\so(10)$ GUT phase (in the presence of the WZW term), as indicated by the DQC region in \figref{fig:DQC}. The DQC region (including the DQC point) is described by the QED$'_4$ theory.

Starting from the DQC phase, condensing the $\Phi_\mathbf{45}$ or $\Phi_\mathbf{54}$ Higgs fields will generate bilinear masses for the fermionic partons and break the $\Spin(10)$ symmetry at the same time. When the fermionic partons become fully gapped, the emergent $\U(1)'$ dark photons can remain deconfined {and gapless}, as the Maxwell theory has a stable deconfined phase in 4d. Therefore, the resulting phases are PS, GG, or SM theories coexisting with a decoupled pure $\U(1)'$ gauge theory. They are denoted as PS$^*$, GG$^*$ and SM$^*$ in \figref{fig:DQC} (where the superscript $*$ indicates the coexistence of dark photons in the low-energy spectrum). 

However if the $\U(1)'$ gauge coupling is strong enough, the $\U(1)'$ gauge theory can also be driven to its confined phase. The confinement transition is indicated by the dashed circle in \figref{fig:DQC}. After $\U(1)'$ confinement, the DQC physics will disappear. The PS$^*$, GG$^*$ and SM$^*$ phases will fall back to the conventional PS, GG, and SM phases. {But the $\so(10)$ GUT phase with the WZW term (even after the $\U(1)'$ confinement) will remain non-trivial due to the $w_2w_3$ anomaly. The anomaly may also be saturated by other CFT (differed from the QED$'_4$ CFT proposed here), or by some TQFT.}

\begin{figure}[htbp]
\begin{center}
\includegraphics[width=0.95\columnwidth]{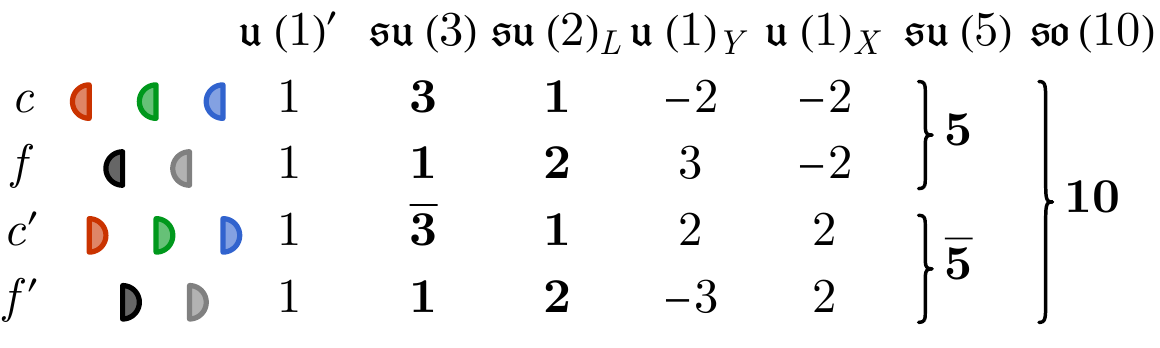}
\caption{Quantum numbers of fermionic partons in the DQC phase.}
\label{fig:parton}
\end{center}
\end{figure}

The quantum numbers of the ten Dirac fermionic partons are summarized in \figref{fig:parton}. In terms of the $\g_\text{SM}$ representations, the fermionic partons have charge assignments that are very different from those of the matter fermions in the SM. Some fermionic partons (call colorons $c,c'$) carries the $\su(3)$ color charge but not the weak $\su(2)_L$ flavor charge, while other fermionic partons (called flavorons $f,f'$) carries the weak $\su(2)_L$ flavor charge but not the color $\su(3)$ charge. This color-flavor separation on the fermionic partons is one important feature of these additional fermions that are fractionalized from the GUT Higgs field. This is analogous to the spin-charge separation \cite{Tomonaga1950Remarks,Luttinger1963An-Exactly,Anderson2000Spin-charge} in condensed matter physics.

Suppose the DQC phase once existed in the early history of our universe, now we have entered the SM$^*$ phase, where the GUT Higgs field $\Phi$ has condensed. All these fermionic partons will be gapped by the Higgs condensation at the GUT scale. {The $\U(1)'$ dark photons are decoupled from the fermionic matter particles in the SM}, and remain as dark cosmic background radiation, which could provide a potential dark matter candidate.

\section{Summary}

Standard lore ritualizes our quantum vacuum in the 4-dimensional spacetime governed by the Standard Models, while lifting towards one of Grand Unifications (GUTs) at higher energy scales. In contrast, in recent studies \cite{Wang2021Gauge1,Wang2021Gauge2,Wang2021Cobordism}, we introduce an alternative view that the SM is a low-energy quantum vacuum arising from various neighbor GUT vacua competition in an immense quantum phase diagram. We study the quantum phase transition between the SM and various GUTs, including the Georgi-Glashow model, the Pati-Salam model, and the $\so(10)$ GUT. We found that the competition between the GG and PS phases could lead to a gapless quantum critical phase, similar to the deconfined quantum critical point in quantum magnets of condensed matter, where the GUT Higgs field fractionalizes into {fermionic} partons coupled by an emergent gauge field. This fractionalization is entailed by a $w_2w_3(TM) = w_2w_3(V_{\SO(10)})$ nonperturbative global anomaly in the presence of a Wess-Zumino-Witten-like term for the GUT Higgs field. The quantum critical phase features fragmentary low-energy excitations of Color-Flavor separation of fermionic partons
and emergent Dark Gauge force, which may leave observable consequences even away from the critical phase.

\acknowledgements{The authors are honored to contribute to
the Professor Chen-Ning Yang Centenary Festschrift. A related presentation
is given by Yi-Zhuang You at the Ultra-Quantum Matter Annual Meeting
on January 21, 2022 \cite{UQMmeeting}.}

\bibliography{ref}

\begin{thebibliography}{57}%
\makeatletter
\providecommand \@ifxundefined [1]{%
 \@ifx{#1\undefined}
}%
\providecommand \@ifnum [1]{%
 \ifnum #1\expandafter \@firstoftwo
 \else \expandafter \@secondoftwo
 \fi
}%
\providecommand \@ifx [1]{%
 \ifx #1\expandafter \@firstoftwo
 \else \expandafter \@secondoftwo
 \fi
}%
\providecommand \natexlab [1]{#1}%
\providecommand \enquote  [1]{``#1''}%
\providecommand \bibnamefont  [1]{#1}%
\providecommand \bibfnamefont [1]{#1}%
\providecommand \citenamefont [1]{#1}%
\providecommand \href@noop [0]{\@secondoftwo}%
\providecommand \href [0]{\begingroup \@sanitize@url \@href}%
\providecommand \@href[1]{\@@startlink{#1}\@@href}%
\providecommand \@@href[1]{\endgroup#1\@@endlink}%
\providecommand \@sanitize@url [0]{\catcode `\\12\catcode `\$12\catcode
  `\&12\catcode `\#12\catcode `\^12\catcode `\_12\catcode `\%12\relax}%
\providecommand \@@startlink[1]{}%
\providecommand \@@endlink[0]{}%
\providecommand \url  [0]{\begingroup\@sanitize@url \@url }%
\providecommand \@url [1]{\endgroup\@href {#1}{\urlprefix }}%
\providecommand \urlprefix  [0]{URL }%
\providecommand \Eprint [0]{\href }%
\providecommand \doibase [0]{https://doi.org/}%
\providecommand \selectlanguage [0]{\@gobble}%
\providecommand \bibinfo  [0]{\@secondoftwo}%
\providecommand \bibfield  [0]{\@secondoftwo}%
\providecommand \translation [1]{[#1]}%
\providecommand \BibitemOpen [0]{}%
\providecommand \bibitemStop [0]{}%
\providecommand \bibitemNoStop [0]{.\EOS\space}%
\providecommand \EOS [0]{\spacefactor3000\relax}%
\providecommand \BibitemShut  [1]{\csname bibitem#1\endcsname}%
\let\auto@bib@innerbib\@empty
\bibitem [{\citenamefont
  {Glashow}(1961)}]{Glashow1961trPartialSymmetriesofWeakInteractions}%
  \BibitemOpen
  \bibfield  {author} {\bibinfo {author} {\bibfnamefont {S.~L.}\ \bibnamefont
  {Glashow}},\ }\bibfield  {title} {\bibinfo {title} {{Partial Symmetries of
  Weak Interactions}},\ }\href {https://doi.org/10.1016/0029-5582(61)90469-2}
  {\bibfield  {journal} {\bibinfo  {journal} {Nucl. Phys.}\ }\textbf {\bibinfo
  {volume} {22}},\ \bibinfo {pages} {579} (\bibinfo {year} {1961})}\BibitemShut
  {NoStop}%
\bibitem [{\citenamefont {Salam}\ and\ \citenamefont
  {Ward}(1964)}]{Salam1964ryElectromagneticWeakInteractions}%
  \BibitemOpen
  \bibfield  {author} {\bibinfo {author} {\bibfnamefont {A.}~\bibnamefont
  {Salam}}\ and\ \bibinfo {author} {\bibfnamefont {J.~C.}\ \bibnamefont
  {Ward}},\ }\bibfield  {title} {\bibinfo {title} {{Electromagnetic and weak
  interactions}},\ }\href {https://doi.org/10.1016/0031-9163(64)90711-5}
  {\bibfield  {journal} {\bibinfo  {journal} {Phys. Lett.}\ }\textbf {\bibinfo
  {volume} {13}},\ \bibinfo {pages} {168} (\bibinfo {year} {1964})}\BibitemShut
  {NoStop}%
\bibitem [{\citenamefont {Salam}(1968)}]{Salam1968}%
  \BibitemOpen
  \bibfield  {author} {\bibinfo {author} {\bibfnamefont {A.}~\bibnamefont
  {Salam}},\ }\bibfield  {title} {\bibinfo {title} {{Weak and Electromagnetic
  Interactions}},\ }\href {https://doi.org/10.1142/9789812795915_0034}
  {\bibfield  {journal} {\bibinfo  {journal} {Conf. Proc. C}\ }\textbf
  {\bibinfo {volume} {680519}},\ \bibinfo {pages} {367} (\bibinfo {year}
  {1968})}\BibitemShut {NoStop}%
\bibitem [{\citenamefont {Weinberg}(1967)}]{Weinberg1967tqSMAModelofLeptons}%
  \BibitemOpen
  \bibfield  {author} {\bibinfo {author} {\bibfnamefont {S.}~\bibnamefont
  {Weinberg}},\ }\bibfield  {title} {\bibinfo {title} {{A Model of Leptons}},\
  }\href {https://doi.org/10.1103/PhysRevLett.19.1264} {\bibfield  {journal}
  {\bibinfo  {journal} {Phys. Rev. Lett.}\ }\textbf {\bibinfo {volume} {19}},\
  \bibinfo {pages} {1264} (\bibinfo {year} {1967})}\BibitemShut {NoStop}%
\bibitem [{\citenamefont {Pati}\ and\ \citenamefont
  {Salam}(1974)}]{Pati1974yyPatiSalamLeptonNumberastheFourthColor}%
  \BibitemOpen
  \bibfield  {author} {\bibinfo {author} {\bibfnamefont {J.~C.}\ \bibnamefont
  {Pati}}\ and\ \bibinfo {author} {\bibfnamefont {A.}~\bibnamefont {Salam}},\
  }\bibfield  {title} {\bibinfo {title} {{Lepton Number as the Fourth Color}},\
  }\href {https://doi.org/10.1103/PhysRevD.10.275, 10.1103/PhysRevD.11.703.2}
  {\bibfield  {journal} {\bibinfo  {journal} {Phys. Rev.}\ }\textbf {\bibinfo
  {volume} {D10}},\ \bibinfo {pages} {275} (\bibinfo {year} {1974})},\ \bibinfo
  {note} {[Erratum: Phys. Rev.D11,703(1975)]}\BibitemShut {NoStop}%
\bibitem [{\citenamefont {Georgi}\ and\ \citenamefont
  {Glashow}(1974)}]{Georgi1974syUnityofAllElementaryParticleForces}%
  \BibitemOpen
  \bibfield  {author} {\bibinfo {author} {\bibfnamefont {H.}~\bibnamefont
  {Georgi}}\ and\ \bibinfo {author} {\bibfnamefont {S.~L.}\ \bibnamefont
  {Glashow}},\ }\bibfield  {title} {\bibinfo {title} {{Unity of All Elementary
  Particle Forces}},\ }\href {https://doi.org/10.1103/PhysRevLett.32.438}
  {\bibfield  {journal} {\bibinfo  {journal} {Phys. Rev. Lett.}\ }\textbf
  {\bibinfo {volume} {32}},\ \bibinfo {pages} {438} (\bibinfo {year}
  {1974})}\BibitemShut {NoStop}%
\bibitem [{\citenamefont {Fritzsch}\ and\ \citenamefont
  {Minkowski}(1975)}]{Fritzsch1974nnMinkowskiUnifiedInteractionsofLeptonsandHadrons}%
  \BibitemOpen
  \bibfield  {author} {\bibinfo {author} {\bibfnamefont {H.}~\bibnamefont
  {Fritzsch}}\ and\ \bibinfo {author} {\bibfnamefont {P.}~\bibnamefont
  {Minkowski}},\ }\bibfield  {title} {\bibinfo {title} {{Unified Interactions
  of Leptons and Hadrons}},\ }\href
  {https://doi.org/10.1016/0003-4916(75)90211-0} {\bibfield  {journal}
  {\bibinfo  {journal} {Annals Phys.}\ }\textbf {\bibinfo {volume} {93}},\
  \bibinfo {pages} {193} (\bibinfo {year} {1975})}\BibitemShut {NoStop}%
\bibitem [{\citenamefont {{Nath}}\ and\ \citenamefont {{Fileviez
  P{\'e}rez}}(2007)}]{Nath2007Proton}%
  \BibitemOpen
  \bibfield  {author} {\bibinfo {author} {\bibfnamefont {P.}~\bibnamefont
  {{Nath}}}\ and\ \bibinfo {author} {\bibfnamefont {P.}~\bibnamefont {{Fileviez
  P{\'e}rez}}},\ }\bibfield  {title} {\bibinfo {title} {{Proton stability in
  grand unified theories, in strings and in branes}},\ }\href
  {https://doi.org/10.1016/j.physrep.2007.02.010} {\bibfield  {journal}
  {\bibinfo  {journal} {Phys. Rept.}\ }\textbf {\bibinfo {volume} {441}},\
  \bibinfo {pages} {191} (\bibinfo {year} {2007})},\ \Eprint
  {https://arxiv.org/abs/hep-ph/0601023} {arXiv:hep-ph/0601023 [hep-ph]}
  \BibitemShut {NoStop}%
\bibitem [{\citenamefont {Sachdev}(2011)}]{subirsachdev2011book}%
  \BibitemOpen
  \bibfield  {author} {\bibinfo {author} {\bibfnamefont {S.}~\bibnamefont
  {Sachdev}},\ }\href {https://doi.org/10.1017/CBO9780511973765} {\emph
  {\bibinfo {title} {Quantum Phase Transitions}}},\ \bibinfo {edition} {2nd}\
  ed.\ (\bibinfo  {publisher} {Cambridge University Press},\ \bibinfo {year}
  {2011})\BibitemShut {NoStop}%
\bibitem [{\citenamefont {{Wang}}\ and\ \citenamefont
  {{You}}(2021{\natexlab{a}})}]{Wang2021Gauge1}%
  \BibitemOpen
  \bibfield  {author} {\bibinfo {author} {\bibfnamefont {J.}~\bibnamefont
  {{Wang}}}\ and\ \bibinfo {author} {\bibfnamefont {Y.-Z.}\ \bibnamefont
  {{You}}},\ }\bibfield  {title} {\bibinfo {title} {{Gauge Enhanced Quantum
  Criticality Beyond the Standard Model}},\ }\href@noop {} {\bibfield
  {journal} {\bibinfo  {journal} {arXiv e-prints}\ ,\ \bibinfo {eid}
  {arXiv:2106.16248}} (\bibinfo {year} {2021}{\natexlab{a}})},\ \Eprint
  {https://arxiv.org/abs/2106.16248} {arXiv:2106.16248 [hep-th]} \BibitemShut
  {NoStop}%
\bibitem [{\citenamefont {{Wang}}\ and\ \citenamefont
  {{You}}(2021{\natexlab{b}})}]{Wang2021Gauge2}%
  \BibitemOpen
  \bibfield  {author} {\bibinfo {author} {\bibfnamefont {J.}~\bibnamefont
  {{Wang}}}\ and\ \bibinfo {author} {\bibfnamefont {Y.-Z.}\ \bibnamefont
  {{You}}},\ }\bibfield  {title} {\bibinfo {title} {{Gauge Enhanced Quantum
  Criticality Between Grand Unifications: Categorical Higher Symmetry
  Retraction}},\ }\href@noop {} {\bibfield  {journal} {\bibinfo  {journal}
  {arXiv e-prints}\ ,\ \bibinfo {eid} {arXiv:2111.10369}} (\bibinfo {year}
  {2021}{\natexlab{b}})},\ \Eprint {https://arxiv.org/abs/2111.10369}
  {arXiv:2111.10369 [hep-th]} \BibitemShut {NoStop}%
\bibitem [{\citenamefont {{Wang}}\ \emph {et~al.}(2021)\citenamefont {{Wang}},
  \citenamefont {{Wan}},\ and\ \citenamefont {{You}}}]{Wang2021Cobordism}%
  \BibitemOpen
  \bibfield  {author} {\bibinfo {author} {\bibfnamefont {J.}~\bibnamefont
  {{Wang}}}, \bibinfo {author} {\bibfnamefont {Z.}~\bibnamefont {{Wan}}},\ and\
  \bibinfo {author} {\bibfnamefont {Y.-Z.}\ \bibnamefont {{You}}},\ }\bibfield
  {title} {\bibinfo {title} {{Cobordism and Deformation Class of the Standard
  Model}},\ }\href@noop {} {\bibfield  {journal} {\bibinfo  {journal} {arXiv
  e-prints}\ ,\ \bibinfo {eid} {arXiv:2112.14765}} (\bibinfo {year} {2021})},\
  \Eprint {https://arxiv.org/abs/2112.14765} {arXiv:2112.14765 [hep-th]}
  \BibitemShut {NoStop}%
\bibitem [{\citenamefont {{Anderson}}(1972)}]{Anderson1972More}%
  \BibitemOpen
  \bibfield  {author} {\bibinfo {author} {\bibfnamefont {P.~W.}\ \bibnamefont
  {{Anderson}}},\ }\bibfield  {title} {\bibinfo {title} {{More Is Different}},\
  }\href {https://doi.org/10.1126/science.177.4047.393} {\bibfield  {journal}
  {\bibinfo  {journal} {Science}\ }\textbf {\bibinfo {volume} {177}},\ \bibinfo
  {pages} {393} (\bibinfo {year} {1972})}\BibitemShut {NoStop}%
\bibitem [{\citenamefont {Landau}\ and\ \citenamefont
  {Lifschitz}(1958)}]{Landau1958}%
  \BibitemOpen
  \bibfield  {author} {\bibinfo {author} {\bibfnamefont {L.~D.}\ \bibnamefont
  {Landau}}\ and\ \bibinfo {author} {\bibfnamefont {E.~M.}\ \bibnamefont
  {Lifschitz}},\ }\href@noop {} {\emph {\bibinfo {title} {Statistical Physics -
  Course of Theoretical Physics Vol 5}}}\ (\bibinfo  {publisher} {Pergamon},\
  \bibinfo {address} {London},\ \bibinfo {year} {1958})\BibitemShut {NoStop}%
\bibitem [{\citenamefont {Senthil}\ \emph {et~al.}(2004)\citenamefont
  {Senthil}, \citenamefont {Vishwanath}, \citenamefont {Balents}, \citenamefont
  {Sachdev},\ and\ \citenamefont {Fisher}}]{SenthildQCP0311326}%
  \BibitemOpen
  \bibfield  {author} {\bibinfo {author} {\bibfnamefont {T.}~\bibnamefont
  {Senthil}}, \bibinfo {author} {\bibfnamefont {A.}~\bibnamefont {Vishwanath}},
  \bibinfo {author} {\bibfnamefont {L.}~\bibnamefont {Balents}}, \bibinfo
  {author} {\bibfnamefont {S.}~\bibnamefont {Sachdev}},\ and\ \bibinfo {author}
  {\bibfnamefont {M.~P.~A.}\ \bibnamefont {Fisher}},\ }\bibfield  {title}
  {\bibinfo {title} {{Deconfined Quantum Critical Points}},\ }\href
  {https://doi.org/10.1126/science.1091806} {\bibfield  {journal} {\bibinfo
  {journal} {Science}\ }\textbf {\bibinfo {volume} {303}},\ \bibinfo {pages}
  {1490} (\bibinfo {year} {2004})},\ \Eprint
  {https://arxiv.org/abs/cond-mat/0311326} {arXiv:cond-mat/0311326}
  \BibitemShut {NoStop}%
\bibitem [{\citenamefont {{Senthil}}\ \emph {et~al.}(2004)\citenamefont
  {{Senthil}}, \citenamefont {{Balents}}, \citenamefont {{Sachdev}},
  \citenamefont {{Vishwanath}},\ and\ \citenamefont
  {{Fisher}}}]{Senthil2004Quantum}%
  \BibitemOpen
  \bibfield  {author} {\bibinfo {author} {\bibfnamefont {T.}~\bibnamefont
  {{Senthil}}}, \bibinfo {author} {\bibfnamefont {L.}~\bibnamefont
  {{Balents}}}, \bibinfo {author} {\bibfnamefont {S.}~\bibnamefont
  {{Sachdev}}}, \bibinfo {author} {\bibfnamefont {A.}~\bibnamefont
  {{Vishwanath}}},\ and\ \bibinfo {author} {\bibfnamefont {M.~P.~A.}\
  \bibnamefont {{Fisher}}},\ }\bibfield  {title} {\bibinfo {title} {{Quantum
  criticality beyond the Landau-Ginzburg-Wilson paradigm}},\ }\href
  {https://doi.org/10.1103/PhysRevB.70.144407} {\bibfield  {journal} {\bibinfo
  {journal} {\prb}\ }\textbf {\bibinfo {volume} {70}},\ \bibinfo {eid} {144407}
  (\bibinfo {year} {2004})},\ \Eprint {https://arxiv.org/abs/cond-mat/0312617}
  {arXiv:cond-mat/0312617 [cond-mat.str-el]} \BibitemShut {NoStop}%
\bibitem [{\citenamefont {{Motrunich}}\ and\ \citenamefont
  {{Vishwanath}}(2004)}]{Motrunich2004Emergent}%
  \BibitemOpen
  \bibfield  {author} {\bibinfo {author} {\bibfnamefont {O.~I.}\ \bibnamefont
  {{Motrunich}}}\ and\ \bibinfo {author} {\bibfnamefont {A.}~\bibnamefont
  {{Vishwanath}}},\ }\bibfield  {title} {\bibinfo {title} {{Emergent photons
  and transitions in the O(3) sigma model with hedgehog suppression}},\ }\href
  {https://doi.org/10.1103/PhysRevB.70.075104} {\bibfield  {journal} {\bibinfo
  {journal} {\prb}\ }\textbf {\bibinfo {volume} {70}},\ \bibinfo {eid} {075104}
  (\bibinfo {year} {2004})},\ \Eprint {https://arxiv.org/abs/cond-mat/0311222}
  {arXiv:cond-mat/0311222 [cond-mat.str-el]} \BibitemShut {NoStop}%
\bibitem [{\citenamefont {{Levin}}\ and\ \citenamefont
  {{Senthil}}(2004)}]{LevinSenthil0405702}%
  \BibitemOpen
  \bibfield  {author} {\bibinfo {author} {\bibfnamefont {M.}~\bibnamefont
  {{Levin}}}\ and\ \bibinfo {author} {\bibfnamefont {T.}~\bibnamefont
  {{Senthil}}},\ }\bibfield  {title} {\bibinfo {title} {{Deconfined quantum
  criticality and N{\'e}el order via dimer disorder}},\ }\href
  {https://doi.org/10.1103/PhysRevB.70.220403} {\bibfield  {journal} {\bibinfo
  {journal} {Phys. Rev. B}\ }\textbf {\bibinfo {volume} {70}},\ \bibinfo {eid}
  {220403} (\bibinfo {year} {2004})},\ \Eprint
  {https://arxiv.org/abs/cond-mat/0405702} {arXiv:cond-mat/0405702
  [cond-mat.str-el]} \BibitemShut {NoStop}%
\bibitem [{\citenamefont {Sandvik}(2007)}]{Sandvik2007Evidence}%
  \BibitemOpen
  \bibfield  {author} {\bibinfo {author} {\bibfnamefont {A.~W.}\ \bibnamefont
  {Sandvik}},\ }\bibfield  {title} {\bibinfo {title} {Evidence for deconfined
  quantum criticality in a two-dimensional heisenberg model with four-spin
  interactions},\ }\href {https://doi.org/10.1103/PhysRevLett.98.227202}
  {\bibfield  {journal} {\bibinfo  {journal} {Phys. Rev. Lett.}\ }\textbf
  {\bibinfo {volume} {98}},\ \bibinfo {pages} {227202} (\bibinfo {year}
  {2007})}\BibitemShut {NoStop}%
\bibitem [{\citenamefont {{Sachdev}}\ and\ \citenamefont
  {{Yin}}(2008)}]{Sachdev2008Quantum}%
  \BibitemOpen
  \bibfield  {author} {\bibinfo {author} {\bibfnamefont {S.}~\bibnamefont
  {{Sachdev}}}\ and\ \bibinfo {author} {\bibfnamefont {X.}~\bibnamefont
  {{Yin}}},\ }\bibfield  {title} {\bibinfo {title} {{Quantum phase transitions
  beyond the Landau-Ginzburg paradigm and supersymmetry}},\ }\href@noop {}
  {\bibfield  {journal} {\bibinfo  {journal} {arXiv e-prints}\ ,\ \bibinfo
  {eid} {arXiv:0808.0191}} (\bibinfo {year} {2008})},\ \Eprint
  {https://arxiv.org/abs/0808.0191} {arXiv:0808.0191 [cond-mat.str-el]}
  \BibitemShut {NoStop}%
\bibitem [{\citenamefont {{Sachdev}}(2009)}]{Sachdev2009Exotic}%
  \BibitemOpen
  \bibfield  {author} {\bibinfo {author} {\bibfnamefont {S.}~\bibnamefont
  {{Sachdev}}},\ }\bibfield  {title} {\bibinfo {title} {{Exotic phases and
  quantum phase transitions: model systems and experiments}},\ }\href@noop {}
  {\bibfield  {journal} {\bibinfo  {journal} {arXiv e-prints}\ ,\ \bibinfo
  {eid} {arXiv:0901.4103}} (\bibinfo {year} {2009})},\ \Eprint
  {https://arxiv.org/abs/0901.4103} {arXiv:0901.4103 [cond-mat.str-el]}
  \BibitemShut {NoStop}%
\bibitem [{\citenamefont {Vishwanath}\ and\ \citenamefont
  {Senthil}(2013)}]{Vishwanath2013Physics}%
  \BibitemOpen
  \bibfield  {author} {\bibinfo {author} {\bibfnamefont {A.}~\bibnamefont
  {Vishwanath}}\ and\ \bibinfo {author} {\bibfnamefont {T.}~\bibnamefont
  {Senthil}},\ }\bibfield  {title} {\bibinfo {title} {Physics of
  three-dimensional bosonic topological insulators: Surface-deconfined
  criticality and quantized magnetoelectric effect},\ }\href
  {https://doi.org/10.1103/PhysRevX.3.011016} {\bibfield  {journal} {\bibinfo
  {journal} {Phys. Rev. X}\ }\textbf {\bibinfo {volume} {3}},\ \bibinfo {pages}
  {011016} (\bibinfo {year} {2013})}\BibitemShut {NoStop}%
\bibitem [{\citenamefont {Yang}\ and\ \citenamefont
  {Mills}(1954)}]{Yang1954Conservation}%
  \BibitemOpen
  \bibfield  {author} {\bibinfo {author} {\bibfnamefont {C.~N.}\ \bibnamefont
  {Yang}}\ and\ \bibinfo {author} {\bibfnamefont {R.~L.}\ \bibnamefont
  {Mills}},\ }\bibfield  {title} {\bibinfo {title} {{Conservation of Isotopic
  Spin and Isotopic Gauge Invariance}},\ }\href
  {https://doi.org/10.1103/PhysRev.96.191} {\bibfield  {journal} {\bibinfo
  {journal} {Phys. Rev.}\ }\textbf {\bibinfo {volume} {96}},\ \bibinfo {pages}
  {191} (\bibinfo {year} {1954})}\BibitemShut {NoStop}%
\bibitem [{\citenamefont {{Drewes}}(2013)}]{Drewes2013The-Phenomenology}%
  \BibitemOpen
  \bibfield  {author} {\bibinfo {author} {\bibfnamefont {M.}~\bibnamefont
  {{Drewes}}},\ }\bibfield  {title} {\bibinfo {title} {{The Phenomenology of
  Right Handed Neutrinos}},\ }\href {https://doi.org/10.1142/S0218301313300191}
  {\bibfield  {journal} {\bibinfo  {journal} {International Journal of Modern
  Physics E}\ }\textbf {\bibinfo {volume} {22}},\ \bibinfo {eid} {1330019-593}
  (\bibinfo {year} {2013})},\ \Eprint {https://arxiv.org/abs/1303.6912}
  {arXiv:1303.6912 [hep-ph]} \BibitemShut {NoStop}%
\bibitem [{\citenamefont {{Boyarsky}}\ \emph {et~al.}(2019)\citenamefont
  {{Boyarsky}}, \citenamefont {{Drewes}}, \citenamefont {{Lasserre}},
  \citenamefont {{Mertens}},\ and\ \citenamefont
  {{Ruchayskiy}}}]{Boyarsky2019Sterile}%
  \BibitemOpen
  \bibfield  {author} {\bibinfo {author} {\bibfnamefont {A.}~\bibnamefont
  {{Boyarsky}}}, \bibinfo {author} {\bibfnamefont {M.}~\bibnamefont
  {{Drewes}}}, \bibinfo {author} {\bibfnamefont {T.}~\bibnamefont
  {{Lasserre}}}, \bibinfo {author} {\bibfnamefont {S.}~\bibnamefont
  {{Mertens}}},\ and\ \bibinfo {author} {\bibfnamefont {O.}~\bibnamefont
  {{Ruchayskiy}}},\ }\bibfield  {title} {\bibinfo {title} {{Sterile neutrino
  Dark Matter}},\ }\href {https://doi.org/10.1016/j.ppnp.2018.07.004}
  {\bibfield  {journal} {\bibinfo  {journal} {Progress in Particle and Nuclear
  Physics}\ }\textbf {\bibinfo {volume} {104}},\ \bibinfo {pages} {1} (\bibinfo
  {year} {2019})},\ \Eprint {https://arxiv.org/abs/1807.07938}
  {arXiv:1807.07938 [hep-ph]} \BibitemShut {NoStop}%
\bibitem [{\citenamefont {Wen}(2013)}]{Wen2013ppa1305.1045}%
  \BibitemOpen
  \bibfield  {author} {\bibinfo {author} {\bibfnamefont {X.-G.}\ \bibnamefont
  {Wen}},\ }\bibfield  {title} {\bibinfo {title} {{A lattice non-perturbative
  definition of an SO(10) chiral gauge theory and its induced standard
  model}},\ }\href {https://doi.org/10.1088/0256-307X/30/11/111101} {\bibfield
  {journal} {\bibinfo  {journal} {Chin. Phys. Lett.}\ }\textbf {\bibinfo
  {volume} {30}},\ \bibinfo {pages} {111101} (\bibinfo {year} {2013})},\
  \Eprint {https://arxiv.org/abs/1305.1045} {arXiv:1305.1045 [hep-lat]}
  \BibitemShut {NoStop}%
\bibitem [{\citenamefont {You}\ and\ \citenamefont
  {Xu}(2015)}]{You2015Interacting}%
  \BibitemOpen
  \bibfield  {author} {\bibinfo {author} {\bibfnamefont {Y.-Z.}\ \bibnamefont
  {You}}\ and\ \bibinfo {author} {\bibfnamefont {C.}~\bibnamefont {Xu}},\
  }\bibfield  {title} {\bibinfo {title} {Interacting topological insulator and
  emergent grand unified theory},\ }\href
  {https://doi.org/10.1103/physrevb.91.125147} {\bibfield  {journal} {\bibinfo
  {journal} {Phys. Rev. B}\ }\textbf {\bibinfo {volume} {91}},\ \bibinfo
  {pages} {125147} (\bibinfo {year} {2015})},\ \Eprint
  {https://arxiv.org/abs/1412.4784} {arXiv:1412.4784} \BibitemShut {NoStop}%
\bibitem [{\citenamefont {BenTov}\ and\ \citenamefont
  {Zee}(2016)}]{BenTov2015graZee1505.04312}%
  \BibitemOpen
  \bibfield  {author} {\bibinfo {author} {\bibfnamefont {Y.}~\bibnamefont
  {BenTov}}\ and\ \bibinfo {author} {\bibfnamefont {A.}~\bibnamefont {Zee}},\
  }\bibfield  {title} {\bibinfo {title} {{Origin of families and $SO(18)$ grand
  unification}},\ }\href {https://doi.org/10.1103/PhysRevD.93.065036}
  {\bibfield  {journal} {\bibinfo  {journal} {Phys. Rev.}\ }\textbf {\bibinfo
  {volume} {D93}},\ \bibinfo {pages} {065036} (\bibinfo {year} {2016})},\
  \Eprint {https://arxiv.org/abs/1505.04312} {arXiv:1505.04312 [hep-th]}
  \BibitemShut {NoStop}%
\bibitem [{\citenamefont {Garcia-Etxebarria}\ and\ \citenamefont
  {Montero}(2019)}]{GarciaEtxebarriaMontero2018ajm1808.00009}%
  \BibitemOpen
  \bibfield  {author} {\bibinfo {author} {\bibfnamefont {I.}~\bibnamefont
  {Garcia-Etxebarria}}\ and\ \bibinfo {author} {\bibfnamefont {M.}~\bibnamefont
  {Montero}},\ }\bibfield  {title} {\bibinfo {title} {{Dai-Freed anomalies in
  particle physics}},\ }\href {https://doi.org/10.1007/JHEP08(2019)003}
  {\bibfield  {journal} {\bibinfo  {journal} {JHEP}\ }\textbf {\bibinfo
  {volume} {08}},\ \bibinfo {pages} {003}},\ \Eprint
  {https://arxiv.org/abs/1808.00009} {arXiv:1808.00009 [hep-th]} \BibitemShut
  {NoStop}%
\bibitem [{\citenamefont {Wan}\ and\ \citenamefont
  {Wang}(2020)}]{WW2019fxh1910.14668}%
  \BibitemOpen
  \bibfield  {author} {\bibinfo {author} {\bibfnamefont {Z.}~\bibnamefont
  {Wan}}\ and\ \bibinfo {author} {\bibfnamefont {J.}~\bibnamefont {Wang}},\
  }\bibfield  {title} {\bibinfo {title} {{Beyond Standard Models and Grand
  Unifications: Anomalies, Topological Terms, and Dynamical Constraints via
  Cobordisms}},\ }\href {https://doi.org/10.1007/JHEP07(2020)062} {\bibfield
  {journal} {\bibinfo  {journal} {JHEP}\ }\textbf {\bibinfo {volume} {07}},\
  \bibinfo {pages} {062}},\ \Eprint {https://arxiv.org/abs/1910.14668}
  {arXiv:1910.14668 [hep-th]} \BibitemShut {NoStop}%
\bibitem [{\citenamefont {{Wang}}(2020)}]{Wang2020Anomaly}%
  \BibitemOpen
  \bibfield  {author} {\bibinfo {author} {\bibfnamefont {J.}~\bibnamefont
  {{Wang}}},\ }\bibfield  {title} {\bibinfo {title} {{Anomaly and Cobordism
  Constraints Beyond the Standard Model: Topological Force}},\ }\href@noop {}
  {\bibfield  {journal} {\bibinfo  {journal} {arXiv e-prints}\ ,\ \bibinfo
  {eid} {arXiv:2006.16996}} (\bibinfo {year} {2020})},\ \Eprint
  {https://arxiv.org/abs/2006.16996} {arXiv:2006.16996 [hep-th]} \BibitemShut
  {NoStop}%
\bibitem [{\citenamefont {Wang}(2020)}]{JW2008.06499}%
  \BibitemOpen
  \bibfield  {author} {\bibinfo {author} {\bibfnamefont {J.}~\bibnamefont
  {Wang}},\ }\bibfield  {title} {\bibinfo {title} {{Anomaly and Cobordism
  Constraints Beyond Grand Unification: Energy Hierarchy}},\ }\href@noop {} {\
  (\bibinfo {year} {2020})},\ \Eprint {https://arxiv.org/abs/2008.06499}
  {arXiv:2008.06499 [hep-th]} \BibitemShut {NoStop}%
\bibitem [{\citenamefont {Wang}(2021)}]{JW2012.15860}%
  \BibitemOpen
  \bibfield  {author} {\bibinfo {author} {\bibfnamefont {J.}~\bibnamefont
  {Wang}},\ }\bibfield  {title} {\bibinfo {title} {{Ultra Unification}},\
  }\href {https://doi.org/10.1103/PhysRevD.103.105024} {\bibfield  {journal}
  {\bibinfo  {journal} {Phys. Rev. D}\ }\textbf {\bibinfo {volume} {103}},\
  \bibinfo {pages} {105024} (\bibinfo {year} {2021})},\ \Eprint
  {https://arxiv.org/abs/2012.15860} {arXiv:2012.15860 [hep-th]} \BibitemShut
  {NoStop}%
\bibitem [{\citenamefont {Wilczek}\ and\ \citenamefont
  {Zee}(1979)}]{Wilczek1979hcZee}%
  \BibitemOpen
  \bibfield  {author} {\bibinfo {author} {\bibfnamefont {F.}~\bibnamefont
  {Wilczek}}\ and\ \bibinfo {author} {\bibfnamefont {A.}~\bibnamefont {Zee}},\
  }\bibfield  {title} {\bibinfo {title} {{Operator Analysis of Nucleon
  Decay}},\ }\href {https://doi.org/10.1103/PhysRevLett.43.1571} {\bibfield
  {journal} {\bibinfo  {journal} {Phys. Rev. Lett.}\ }\textbf {\bibinfo
  {volume} {43}},\ \bibinfo {pages} {1571} (\bibinfo {year}
  {1979})}\BibitemShut {NoStop}%
\bibitem [{\citenamefont {Baez}\ and\ \citenamefont
  {Huerta}(2010)}]{BaezHuerta0904.1556}%
  \BibitemOpen
  \bibfield  {author} {\bibinfo {author} {\bibfnamefont {J.~C.}\ \bibnamefont
  {Baez}}\ and\ \bibinfo {author} {\bibfnamefont {J.}~\bibnamefont {Huerta}},\
  }\bibfield  {title} {\bibinfo {title} {{The Algebra of Grand Unified
  Theories}},\ }\href {https://doi.org/10.1090/S0273-0979-10-01294-2}
  {\bibfield  {journal} {\bibinfo  {journal} {Bull. Am. Math. Soc.}\ }\textbf
  {\bibinfo {volume} {47}},\ \bibinfo {pages} {483} (\bibinfo {year} {2010})},\
  \Eprint {https://arxiv.org/abs/0904.1556} {arXiv:0904.1556 [hep-th]}
  \BibitemShut {NoStop}%
\bibitem [{\citenamefont {Fidkowski}\ and\ \citenamefont
  {Kitaev}(2010)}]{FidkowskifSPT2}%
  \BibitemOpen
  \bibfield  {author} {\bibinfo {author} {\bibfnamefont {L.}~\bibnamefont
  {Fidkowski}}\ and\ \bibinfo {author} {\bibfnamefont {A.}~\bibnamefont
  {Kitaev}},\ }\bibfield  {title} {\bibinfo {title} {Effects of interactions on
  the topological classification of free fermion systems},\ }\href
  {https://doi.org/10.1103/PhysRevB.81.134509} {\bibfield  {journal} {\bibinfo
  {journal} {Phys. Rev. B}\ }\textbf {\bibinfo {volume} {81}},\ \bibinfo
  {pages} {134509} (\bibinfo {year} {2010})}\BibitemShut {NoStop}%
\bibitem [{\citenamefont {Wang}\ and\ \citenamefont
  {Wen}(2013)}]{Wang2013ytaJW1307.7480}%
  \BibitemOpen
  \bibfield  {author} {\bibinfo {author} {\bibfnamefont {J.}~\bibnamefont
  {Wang}}\ and\ \bibinfo {author} {\bibfnamefont {X.-G.}\ \bibnamefont {Wen}},\
  }\bibfield  {title} {\bibinfo {title} {{Non-Perturbative Regularization of
  1+1D Anomaly-Free Chiral Fermions and Bosons: On the equivalence of anomaly
  matching conditions and boundary gapping rules}},\ }\href@noop {} {\
  (\bibinfo {year} {2013})},\ \Eprint {https://arxiv.org/abs/1307.7480}
  {arXiv:1307.7480 [hep-lat]} \BibitemShut {NoStop}%
\bibitem [{\citenamefont {Wang}\ and\ \citenamefont
  {Wen}(2019)}]{Wang2018ugfJW1807.05998}%
  \BibitemOpen
  \bibfield  {author} {\bibinfo {author} {\bibfnamefont {J.}~\bibnamefont
  {Wang}}\ and\ \bibinfo {author} {\bibfnamefont {X.-G.}\ \bibnamefont {Wen}},\
  }\bibfield  {title} {\bibinfo {title} {{A Solution to the 1+1D Gauged Chiral
  Fermion Problem}},\ }\href {https://doi.org/10.1103/PhysRevD.99.111501}
  {\bibfield  {journal} {\bibinfo  {journal} {Phys. Rev.}\ }\textbf {\bibinfo
  {volume} {D99}},\ \bibinfo {pages} {111501} (\bibinfo {year} {2019})},\
  \Eprint {https://arxiv.org/abs/1807.05998} {arXiv:1807.05998 [hep-lat]}
  \BibitemShut {NoStop}%
\bibitem [{\citenamefont {You}\ \emph {et~al.}(2018{\natexlab{a}})\citenamefont
  {You}, \citenamefont {He}, \citenamefont {Xu},\ and\ \citenamefont
  {Vishwanath}}]{YouHeXuVishwanath1705.09313}%
  \BibitemOpen
  \bibfield  {author} {\bibinfo {author} {\bibfnamefont {Y.-Z.}\ \bibnamefont
  {You}}, \bibinfo {author} {\bibfnamefont {Y.-C.}\ \bibnamefont {He}},
  \bibinfo {author} {\bibfnamefont {C.}~\bibnamefont {Xu}},\ and\ \bibinfo
  {author} {\bibfnamefont {A.}~\bibnamefont {Vishwanath}},\ }\bibfield  {title}
  {\bibinfo {title} {{Symmetric Fermion Mass Generation as Deconfined Quantum
  Criticality}},\ }\href {https://doi.org/10.1103/PhysRevX.8.011026} {\bibfield
   {journal} {\bibinfo  {journal} {Phys. Rev. X}\ }\textbf {\bibinfo {volume}
  {8}},\ \bibinfo {pages} {011026} (\bibinfo {year} {2018}{\natexlab{a}})},\
  \Eprint {https://arxiv.org/abs/1705.09313} {arXiv:1705.09313
  [cond-mat.str-el]} \BibitemShut {NoStop}%
\bibitem [{\citenamefont {You}\ \emph {et~al.}(2018{\natexlab{b}})\citenamefont
  {You}, \citenamefont {He}, \citenamefont {Vishwanath},\ and\ \citenamefont
  {Xu}}]{YouHeVishwanathXu1711.00863}%
  \BibitemOpen
  \bibfield  {author} {\bibinfo {author} {\bibfnamefont {Y.-Z.}\ \bibnamefont
  {You}}, \bibinfo {author} {\bibfnamefont {Y.-C.}\ \bibnamefont {He}},
  \bibinfo {author} {\bibfnamefont {A.}~\bibnamefont {Vishwanath}},\ and\
  \bibinfo {author} {\bibfnamefont {C.}~\bibnamefont {Xu}},\ }\bibfield
  {title} {\bibinfo {title} {{From Bosonic Topological Transition to Symmetric
  Fermion Mass Generation}},\ }\href
  {https://doi.org/10.1103/PhysRevB.97.125112} {\bibfield  {journal} {\bibinfo
  {journal} {Phys. Rev. B}\ }\textbf {\bibinfo {volume} {97}},\ \bibinfo
  {pages} {125112} (\bibinfo {year} {2018}{\natexlab{b}})},\ \Eprint
  {https://arxiv.org/abs/1711.00863} {arXiv:1711.00863 [cond-mat.str-el]}
  \BibitemShut {NoStop}%
\bibitem [{\citenamefont {Eichten}\ and\ \citenamefont
  {Preskill}(1986)}]{Eichten1985ftPreskill1986}%
  \BibitemOpen
  \bibfield  {author} {\bibinfo {author} {\bibfnamefont {E.}~\bibnamefont
  {Eichten}}\ and\ \bibinfo {author} {\bibfnamefont {J.}~\bibnamefont
  {Preskill}},\ }\bibfield  {title} {\bibinfo {title} {{Chiral Gauge Theories
  on the Lattice}},\ }\href {https://doi.org/10.1016/0550-3213(86)90207-5}
  {\bibfield  {journal} {\bibinfo  {journal} {Nucl. Phys.}\ }\textbf {\bibinfo
  {volume} {B268}},\ \bibinfo {pages} {179} (\bibinfo {year}
  {1986})}\BibitemShut {NoStop}%
\bibitem [{\citenamefont {Kikukawa}(2019)}]{Kikukawa2017ngf1710.11618}%
  \BibitemOpen
  \bibfield  {author} {\bibinfo {author} {\bibfnamefont {Y.}~\bibnamefont
  {Kikukawa}},\ }\bibfield  {title} {\bibinfo {title} {{On the gauge invariant
  path-integral measure for the overlap Weyl fermions in $\underline{16}$ of
  SO(10)}},\ }\href {https://doi.org/10.1093/ptep/ptz115} {\bibfield  {journal}
  {\bibinfo  {journal} {PTEP}\ }\textbf {\bibinfo {volume} {2019}},\ \bibinfo
  {pages} {113B03} (\bibinfo {year} {2019})},\ \Eprint
  {https://arxiv.org/abs/1710.11618} {arXiv:1710.11618 [hep-lat]} \BibitemShut
  {NoStop}%
\bibitem [{\citenamefont {Catterall}(2020)}]{Catterall2020fep}%
  \BibitemOpen
  \bibfield  {author} {\bibinfo {author} {\bibfnamefont {S.}~\bibnamefont
  {Catterall}},\ }\bibfield  {title} {\bibinfo {title} {{Chiral Lattice
  Theories From Staggered Fermions}},\ }\href@noop {} {\  (\bibinfo {year}
  {2020})},\ \Eprint {https://arxiv.org/abs/2010.02290} {arXiv:2010.02290
  [hep-lat]} \BibitemShut {NoStop}%
\bibitem [{\citenamefont {Catterall}\ \emph {et~al.}(2021)\citenamefont
  {Catterall}, \citenamefont {Toga},\ and\ \citenamefont
  {Butt}}]{CatterallTogaButt2101.01026}%
  \BibitemOpen
  \bibfield  {author} {\bibinfo {author} {\bibfnamefont {S.}~\bibnamefont
  {Catterall}}, \bibinfo {author} {\bibfnamefont {G.~C.}\ \bibnamefont
  {Toga}},\ and\ \bibinfo {author} {\bibfnamefont {N.}~\bibnamefont {Butt}},\
  }\bibfield  {title} {\bibinfo {title} {{Symmetric mass generation for
  K\"ahler-Dirac fermions}},\ }\href@noop {} {\  (\bibinfo {year} {2021})},\
  \Eprint {https://arxiv.org/abs/2101.01026} {arXiv:2101.01026 [hep-th]}
  \BibitemShut {NoStop}%
\bibitem [{\citenamefont {Razamat}\ and\ \citenamefont
  {Tong}(2021)}]{RazamatTong2009.05037}%
  \BibitemOpen
  \bibfield  {author} {\bibinfo {author} {\bibfnamefont {S.~S.}\ \bibnamefont
  {Razamat}}\ and\ \bibinfo {author} {\bibfnamefont {D.}~\bibnamefont {Tong}},\
  }\bibfield  {title} {\bibinfo {title} {{Gapped Chiral Fermions}},\ }\href
  {https://doi.org/10.1103/PhysRevX.11.011063} {\bibfield  {journal} {\bibinfo
  {journal} {Phys. Rev. X}\ }\textbf {\bibinfo {volume} {11}},\ \bibinfo
  {pages} {011063} (\bibinfo {year} {2021})},\ \Eprint
  {https://arxiv.org/abs/2009.05037} {arXiv:2009.05037 [hep-th]} \BibitemShut
  {NoStop}%
\bibitem [{\citenamefont {Tong}(2021)}]{Tong2104.03997}%
  \BibitemOpen
  \bibfield  {author} {\bibinfo {author} {\bibfnamefont {D.}~\bibnamefont
  {Tong}},\ }\bibfield  {title} {\bibinfo {title} {{Comments on Symmetric Mass
  Generation in 2d and 4d}},\ }\href@noop {} {\  (\bibinfo {year} {2021})},\
  \Eprint {https://arxiv.org/abs/2104.03997} {arXiv:2104.03997 [hep-th]}
  \BibitemShut {NoStop}%
\bibitem [{\citenamefont {Putrov}\ \emph {et~al.}(2017)\citenamefont {Putrov},
  \citenamefont {Wang},\ and\ \citenamefont
  {Yau}}]{Putrov2016qdo1612.09298PWY}%
  \BibitemOpen
  \bibfield  {author} {\bibinfo {author} {\bibfnamefont {P.}~\bibnamefont
  {Putrov}}, \bibinfo {author} {\bibfnamefont {J.}~\bibnamefont {Wang}},\ and\
  \bibinfo {author} {\bibfnamefont {S.-T.}\ \bibnamefont {Yau}},\ }\bibfield
  {title} {\bibinfo {title} {{Braiding Statistics and Link Invariants of
  Bosonic/Fermionic Topological Quantum Matter in 2+1 and 3+1 dimensions}},\
  }\href {https://doi.org/10.1016/j.aop.2017.06.019} {\bibfield  {journal}
  {\bibinfo  {journal} {Annals Phys.}\ }\textbf {\bibinfo {volume} {384}},\
  \bibinfo {pages} {254} (\bibinfo {year} {2017})},\ \Eprint
  {https://arxiv.org/abs/1612.09298} {arXiv:1612.09298 [cond-mat.str-el]}
  \BibitemShut {NoStop}%
\bibitem [{\citenamefont {Guo}\ \emph {et~al.}(2020)\citenamefont {Guo},
  \citenamefont {Ohmori}, \citenamefont {Putrov}, \citenamefont {Wan},\ and\
  \citenamefont {Wang}}]{GuoJW1812.11959}%
  \BibitemOpen
  \bibfield  {author} {\bibinfo {author} {\bibfnamefont {M.}~\bibnamefont
  {Guo}}, \bibinfo {author} {\bibfnamefont {K.}~\bibnamefont {Ohmori}},
  \bibinfo {author} {\bibfnamefont {P.}~\bibnamefont {Putrov}}, \bibinfo
  {author} {\bibfnamefont {Z.}~\bibnamefont {Wan}},\ and\ \bibinfo {author}
  {\bibfnamefont {J.}~\bibnamefont {Wang}},\ }\bibfield  {title} {\bibinfo
  {title} {{Fermionic Finite-Group Gauge Theories and Interacting
  Symmetric/Crystalline Orders via Cobordisms}},\ }\href
  {https://doi.org/10.1007/s00220-019-03671-6} {\bibfield  {journal} {\bibinfo
  {journal} {Commun. Math. Phys.}\ }\textbf {\bibinfo {volume} {376}},\
  \bibinfo {pages} {1073} (\bibinfo {year} {2020})},\ \Eprint
  {https://arxiv.org/abs/1812.11959} {arXiv:1812.11959 [hep-th]} \BibitemShut
  {NoStop}%
\bibitem [{\citenamefont {Wang}\ and\ \citenamefont
  {Wen}(2020)}]{WangWen2018cai1809.11171}%
  \BibitemOpen
  \bibfield  {author} {\bibinfo {author} {\bibfnamefont {J.}~\bibnamefont
  {Wang}}\ and\ \bibinfo {author} {\bibfnamefont {X.-G.}\ \bibnamefont {Wen}},\
  }\bibfield  {title} {\bibinfo {title} {{A Non-Perturbative Definition of the
  Standard Models}},\ }\href {https://doi.org/10.1103/PhysRevResearch.2.023356}
  {\bibfield  {journal} {\bibinfo  {journal} {Phys. Rev. Res.}\ }\textbf
  {\bibinfo {volume} {2}},\ \bibinfo {pages} {023356} (\bibinfo {year}
  {2020})},\ \Eprint {https://arxiv.org/abs/1809.11171} {arXiv:1809.11171
  [hep-th]} \BibitemShut {NoStop}%
\bibitem [{\citenamefont {Wan}\ and\ \citenamefont
  {Wang}(2019)}]{WanWang2018bns1812.11967}%
  \BibitemOpen
  \bibfield  {author} {\bibinfo {author} {\bibfnamefont {Z.}~\bibnamefont
  {Wan}}\ and\ \bibinfo {author} {\bibfnamefont {J.}~\bibnamefont {Wang}},\
  }\bibfield  {title} {\bibinfo {title} {{Higher Anomalies, Higher Symmetries,
  and Cobordisms I: Classification of Higher-Symmetry-Protected Topological
  States and Their Boundary Fermionic/Bosonic Anomalies via a Generalized
  Cobordism Theory}},\ }\href {https://doi.org/10.4310/AMSA.2019.v4.n2.a2}
  {\bibfield  {journal} {\bibinfo  {journal} {Ann. Math. Sci. Appl.}\ }\textbf
  {\bibinfo {volume} {4}},\ \bibinfo {pages} {107} (\bibinfo {year} {2019})},\
  \Eprint {https://arxiv.org/abs/1812.11967} {arXiv:1812.11967 [hep-th]}
  \BibitemShut {NoStop}%
\bibitem [{\citenamefont {Wang}\ \emph {et~al.}(2019)\citenamefont {Wang},
  \citenamefont {Wen},\ and\ \citenamefont
  {Witten}}]{WangWenWitten2018qoy1810.00844}%
  \BibitemOpen
  \bibfield  {author} {\bibinfo {author} {\bibfnamefont {J.}~\bibnamefont
  {Wang}}, \bibinfo {author} {\bibfnamefont {X.-G.}\ \bibnamefont {Wen}},\ and\
  \bibinfo {author} {\bibfnamefont {E.}~\bibnamefont {Witten}},\ }\bibfield
  {title} {\bibinfo {title} {{A New SU(2) Anomaly}},\ }\href
  {https://doi.org/10.1063/1.5082852} {\bibfield  {journal} {\bibinfo
  {journal} {J. Math. Phys.}\ }\textbf {\bibinfo {volume} {60}},\ \bibinfo
  {pages} {052301} (\bibinfo {year} {2019})},\ \Eprint
  {https://arxiv.org/abs/1810.00844} {arXiv:1810.00844 [hep-th]} \BibitemShut
  {NoStop}%
\bibitem [{\citenamefont {Zhou}\ \emph {et~al.}(2017)\citenamefont {Zhou},
  \citenamefont {Kanoda},\ and\ \citenamefont {Ng}}]{Zhou2017Quantum}%
  \BibitemOpen
  \bibfield  {author} {\bibinfo {author} {\bibfnamefont {Y.}~\bibnamefont
  {Zhou}}, \bibinfo {author} {\bibfnamefont {K.}~\bibnamefont {Kanoda}},\ and\
  \bibinfo {author} {\bibfnamefont {T.-K.}\ \bibnamefont {Ng}},\ }\bibfield
  {title} {\bibinfo {title} {Quantum spin liquid states},\ }\href
  {https://doi.org/10.1103/RevModPhys.89.025003} {\bibfield  {journal}
  {\bibinfo  {journal} {Rev. Mod. Phys.}\ }\textbf {\bibinfo {volume} {89}},\
  \bibinfo {pages} {025003} (\bibinfo {year} {2017})}\BibitemShut {NoStop}%
\bibitem [{\citenamefont {{Savary}}\ and\ \citenamefont
  {{Balents}}(2017)}]{Savary2017Quantum}%
  \BibitemOpen
  \bibfield  {author} {\bibinfo {author} {\bibfnamefont {L.}~\bibnamefont
  {{Savary}}}\ and\ \bibinfo {author} {\bibfnamefont {L.}~\bibnamefont
  {{Balents}}},\ }\bibfield  {title} {\bibinfo {title} {{Quantum spin liquids:
  a review}},\ }\href {https://doi.org/10.1088/0034-4885/80/1/016502}
  {\bibfield  {journal} {\bibinfo  {journal} {Reports on Progress in Physics}\
  }\textbf {\bibinfo {volume} {80}},\ \bibinfo {eid} {016502} (\bibinfo {year}
  {2017})},\ \Eprint {https://arxiv.org/abs/1601.03742} {arXiv:1601.03742
  [cond-mat.str-el]} \BibitemShut {NoStop}%
\bibitem [{\citenamefont {{Tomonaga}}(1950)}]{Tomonaga1950Remarks}%
  \BibitemOpen
  \bibfield  {author} {\bibinfo {author} {\bibfnamefont {S.}~\bibnamefont
  {{Tomonaga}}},\ }\bibfield  {title} {\bibinfo {title} {{Remarks on Bloch's
  Method of Sound Waves applied to Many-Fermion Problems}},\ }\href
  {https://doi.org/10.1143/ptp/5.4.544} {\bibfield  {journal} {\bibinfo
  {journal} {Progress of Theoretical Physics}\ }\textbf {\bibinfo {volume}
  {5}},\ \bibinfo {pages} {544} (\bibinfo {year} {1950})}\BibitemShut {NoStop}%
\bibitem [{\citenamefont {{Luttinger}}(1963)}]{Luttinger1963An-Exactly}%
  \BibitemOpen
  \bibfield  {author} {\bibinfo {author} {\bibfnamefont {J.~M.}\ \bibnamefont
  {{Luttinger}}},\ }\bibfield  {title} {\bibinfo {title} {{An Exactly Soluble
  Model of a Many-Fermion System}},\ }\href {https://doi.org/10.1063/1.1704046}
  {\bibfield  {journal} {\bibinfo  {journal} {Journal of Mathematical Physics}\
  }\textbf {\bibinfo {volume} {4}},\ \bibinfo {pages} {1154} (\bibinfo {year}
  {1963})}\BibitemShut {NoStop}%
\bibitem [{\citenamefont {{Anderson}}(2000)}]{Anderson2000Spin-charge}%
  \BibitemOpen
  \bibfield  {author} {\bibinfo {author} {\bibfnamefont {P.~W.}\ \bibnamefont
  {{Anderson}}},\ }\bibfield  {title} {\bibinfo {title} {{Spin-charge
  separation is the key to the high Tc cuprates}},\ }\href
  {https://doi.org/10.1016/S0921-4534(00)00378-6} {\bibfield  {journal}
  {\bibinfo  {journal} {Physica C Superconductivity}\ }\textbf {\bibinfo
  {volume} {341}},\ \bibinfo {pages} {9} (\bibinfo {year} {2000})},\ \Eprint
  {https://arxiv.org/abs/cond-mat/0007185} {arXiv:cond-mat/0007185
  [cond-mat.str-el]} \BibitemShut {NoStop}%
\bibitem [{UQM(2022)}]{UQMmeeting}%
  \BibitemOpen
  \bibfield  {title} {\bibinfo {title}
  {\href{https://www.simonsfoundation.org/event/simons-collaboration-on-ultra-quantum-matter-annual-meeting-2022/}{https://www.simonsfoundation.org/event/simons-collaboration-on-ultra-quantum-matter-annual-meeting-2022/}},\
  }\href@noop {} {\  (\bibinfo {year} {2022})}\BibitemShut {NoStop}%
\end{thebibliography}%
\end{document}